\documentclass[journal]{IEEEtran}
\usepackage[utf8]{inputenc}
\usepackage{amsmath,amssymb,amsfonts}
\usepackage{algorithm} 
\usepackage{epstopdf}
\usepackage{subcaption} % 使用 subcaption 来处理子图
\usepackage{caption}    % 使用 caption 宏包控制 caption 样式
\usepackage{textcomp}
\usepackage{bm}
\usepackage{amsmath,amsfonts}
\usepackage{algorithmic}
\usepackage{array}
\usepackage{textcomp}
\usepackage{stfloats}
\usepackage{url}
\usepackage{verbatim}
\usepackage{graphicx}
\usepackage{booktabs}
\usepackage{balance}
\usepackage{multirow}
\usepackage{diagbox}
\usepackage{color,xcolor}
\def\BibTeX{{\rm B\kern-.05em{\sc i\kern-.025em b}\kern-.08em
		T\kern-.1667em\lower.7ex\hbox{E}\kern-.125emX}}
\allowdisplaybreaks
\usepackage{xpatch}
\begin{document}

	\title{\huge Aerial RIS-Enhanced Communications: Joint UAV Trajectory, Altitude Control, and Phase Shift Design}
	\author{Bin Li,~Dongdong Yang,~Lei Liu,~and Dusit Niyato,~\IEEEmembership{Fellow,~IEEE}
		\thanks{Bin Li and Dongdong Yang are with the School of Computer Science, Nanjing University of Information Science and Technology, Nanjing 210044, China (e-mail: bin.li@nuist.edu.cn; 202312200024@nuist.edu.cn).}%
		\thanks{Lei Liu is with the Guangzhou Institute of Technology, Xidian University, Guangzhou 510555, China (e-mail: tianjiaoliulei@163.com).}
		\thanks{Dusit Niyato is with the College of Computing and Data Science, Nanyang Technological University, Singapore (e-mail: dniyato@ntu.edu.sg).}}

	\setlength{\parskip}{0pt} % 段落间距
	\maketitle
	
	\begin{abstract}
	Reconfigurable intelligent surface (RIS) has emerged as a pivotal technology for enhancing  wireless networks. Compared to terrestrial RIS deployed on building facades, aerial RIS (ARIS) mounted on quadrotor unmanned aerial vehicle (UAV) offers superior flexibility and extended coverage. However, the inevitable tilt and altitude variations of a quadrotor UAV during flight may lead to severe beam misalignment, significantly degrading ARIS's performance. To address this challenge, we propose an Euler angles-based ARIS control scheme that jointly optimizes the altitude and trajectory of the ARIS by leveraging the UAV's dynamic model. Considering the constraints on ARIS flight energy consumption, flight safety, and the transmission power of a base station (BS), we jointly design the ARIS's altitude, trajectory, phase shifts, and BS beamforming to maximize the system sum‐rate. Due to the continuous control nature of ARIS flight and the strong coupling among variables, we formulate the problem as a Markov decision process and adopt a soft actor-critic algorithm with prioritized experience replay to learn efficient ARIS control policies. Based on the optimized ARIS configuration, we further employ the water-filling and bisection method to efficiently determine the optimal BS beamforming. Numerical results demonstrate that the proposed algorithm significantly outperforms benchmarks in both convergence and communication performance, achieving approximately 14.4\% improvement in sum‐rate. Moreover, in comparison to the fixed-horizontal ARIS scheme, the proposed scheme yields more adaptive trajectories and significantly mitigates performance degradation caused by ARIS tilting, demonstrating strong potential for practical ARIS deployment.
	\end{abstract}
	
	\begin{IEEEkeywords}
	Reconfigurable intelligent surface, UAV altitude, Euler angle, multi-user communication, deep reinforcement learning.
	\end{IEEEkeywords}
	\section{Introduction}
	As a paradigm-shifting wireless communication technology, reconfigurable intelligent surface (RIS) leverages massive low-cost passive elements to achieve programmable signal enhancement via phase-coherent superposition, offering unprecedented advantages in low-power implementation and economical deployment \cite{9424177}. However, conventional terrestrial RIS is constrained by its fixed deployment, limiting service area to static coverage regions \cite{10345491}. This limitation can be mitigated by integrating RIS with unmanned aerial vehicle (UAV), renowned for their superior line-of-sight (LoS) probability and three-dimensional maneuverability \cite{8579209}. The resultant aerial RIS (ARIS) architecture synergistically integrates the complementary benefits of both technologies, establishing itself as a promising solution for next-generation adaptive networks with dynamic beamforming capabilities and extended service coverage \cite{9749767}.
	
	However, in practical ARIS deployments, a UAV inevitably experiences fuselage tilting due to inertial resistance during acceleration/deceleration and aerodynamic effects \cite{8010301}, leading to beam misalignment and channel variations that degrades ARIS-assisted communications \cite{10736523}. Furthermore, existing research has demonstrated that the practical gain of RIS is highly sensitive to signal incidence and reflection angles \cite{9722711}. Despite these physical constraints, current studies predominantly neglect the impact of ARIS altitude variations, resulting in suboptimal system performance that fails to achieve the theoretical upper-bound of ARIS gains \cite{10301685}. This persistent oversight in system modeling fundamentally limits the practical implementation effectiveness of ARIS, presenting a critical challenge remaining to address in ARIS deployment optimization.
	\subsection{Prior Work}
	\subsubsection{RIS-Assisted Communications}
	To fully leverage the channel enhancement benefits of RIS in wireless communications, extensive efforts have been devoted to exploring RIS applications across various communication scenarios. In particular, Guo \textit{et al.} \cite{8982186} explored the application of RIS in a downlink scenario, employing fractional programming and descent-based methods to enhance the sum-rate. Similarly, Yang \textit{et al.} \cite{9497709} addressed resource allocation challenges in a distributed RIS-enabled wireless network and introduced two distinct algorithms tailored for both single-user and multi-user cases. More recently, RIS has also been applied to wireless powered mobile edge computing networks. Zhai \textit{et al.} \cite{10909449} proposed a Stackelberg game-based offloading framework, aiming to enable efficient energy trading and computation between passive devices and the energy station.
	 Considering the half-space coverage limitation of conventional RIS, Xu \textit{et al.} \cite{9437234} proposed the simultaneously transmitting and reflecting RIS (STAR-RIS) architecture, extending its service to full-space domains through its simultaneous transmission and reflection capabilities. In \cite{9570143}, Mu \textit{et al.} investigated STAR-RIS-assisted MISO systems, establishing three fundamental operating protocols and developing a penalty-based iterative algorithm with successive convex approximation.
	Moreover, building on the concept of STAR-RIS, the intelligent omni-surface (IOS) has been proposed in \cite{9722826} which enables simultaneous reflection and refraction to achieve full-dimensional coverage. A hybrid beamforming scheme and prototype validation further demonstrated the feasibility and potential of IOS-assisted communications.
	Driven by the aforementioned advantages of RIS in wireless communications, several studies have explored its role in enhancing UAV-assisted networks, where the UAV functions as an aerial base station (BS). For instance, Li \textit{et al.} \cite{8959174} conducted a joint design of UAV trajectory and RIS passive beamforming to enhance the average achieve rate. Considering the constrained energy capacity of UAV, Liu \textit{et al.} \cite{9277627} proposed a deep Q-network (DQN)-based approach to optimization UAV trajectory and power allocation, aiming to minimize the energy consumption. 
	Furthermore, Zhai \textit{et al.} \cite{10289638} promoted this paradigm to wireless powered communication networks, and proposed a hierarchical Stackelberg game method to address sum-rate and fairness tradeoffs while enhancing utility.
	However, most existing RIS-assisted schemes assume fixed terrestrial deployment, which limits their adaptability to dynamic user distributions and environmental variations. This motivates the integration of UAV and RIS to enhance coverage and flexibility.
	
	\subsubsection{ARIS-Assisted Communications}
	Currently, ARIS trajectory and phase shifts optimization methods generally fall into two main categories, traditional mathematical optimization technologies and data-driven machine learning approaches. For example, Liu \textit{et al.} \cite{liu2022throughput} jointly optimized ARIS trajectory and dynamic power allocation to maximize average downlink throughput in time-slotted transmissions. Furthermore, considering the influence of the incident and reflected angles of signals, Liu \textit{et al.} \cite{10287195} took into account the elevation angle and established an optimization problem with the minimum average achievable rate maximization as the optimization objective, jointly optimizing communication resource allocation, ARIS phase shifts, and trajectory by an efficient iterative algorithm. Deep reinforcement learning (DRL) has become a cornerstone methodology for intelligent aerial network, particularly in joint UAV trajectory and RIS configurations optimization under dynamic channel conditions and operational uncertainties \cite{7792374}. Peng \textit{et al.} \cite{10051712} proposed an energy-harvesting ARIS scheme to enhance UAV endurance and developed a soft-max deep deterministic policy gradient (DDPG)-based solution. To address the massive access demands of GUs, Yao \textit{et al.} \cite{10445520} integrated the ARIS into a satellite-air-ground integrated relay network and proposed an algorithm combining long short-term memory and double DQN to  maximize the system ergodic rate with limited flight energy consumption. Considering the half-space coverage limitation of the RIS, Aung \textit{et al.} \cite{10458888} introduced the aerial STAR-RIS into the mobile edge computing system and utilized a proximal policy optimization (PPO)-based DRL approach to design the UAV trajectory, STAR-RIS configurations, and task offloading strategies. Although ARIS improves coverage and adaptability, existing work primarily focused on trajectory and phase shift optimization while neglecting UAV altitude variations, which may influence the ARIS gain, thereby degrading communication performance.
	
	\subsubsection{RIS Orientation and UAV Tilt}
	Recent studies have demonstrated the significant impact of RIS orientation on overall performance. In \cite{9722711}, Cheng \textit{et al.} systematically quantified the impact of RIS orientation on communications, introducing rotation as an auxiliary control dimension to augment the channel gain of RIS.
	Similarly, in \cite{9201413}, Zeng \textit{et al.} analyzed a downlink RIS-assisted network with one BS and one user, and proposed a coverage maximization algorithm with a closed-form solution for optimal RIS orientation.
	To further enhance the effectiveness of RIS in extending cell coverage, Zeng \textit{et al.} \cite{9201413} examined a downlink RIS-enhanced network comprising single BS and user, and focused on the optimization of RIS orientation and position to enhance overall coverage. Furthermore, in \cite{10444937}, Wang \textit{et al.} explored the rotation of STAR-RIS and utilized deep learning to optimize STAR-RIS orientation in various scenarios, achieving full-space coverage while maximizing STAR-RIS gain. Li \textit{et al.} \cite{10944480} and Yang \textit{et al.} \cite{10891163} studied rotatable RIS-assisted and rotatable STAR-RIS-assisted mobile edge computing systems, respectively.
	
	%Existing research has rigorously demonstrated the critical influence of UAV attitude on UAV-assisted communication performance.
	On the other hand, the impact of UAV tilt on communication performance has also been explored. As a representative study, Wang \textit{et al.} \cite{9573459} systematically investigated UAV jitter effects in millimeter-wave (mmWave) systems and established an unified planar array-based mmWave channel model by analyzing spatial correlations among antenna elements, deriving explicit mathematical relationships between UAV's tilt and 3D positional coordinates. Ouyang \textit{et al.} \cite{10891803} investigated a robust beamforming scheme for rate-splitting multiple access-enabled UAV uplink communication systems under UAV jitter-induced effects, and developed a second-order Taylor series expansion-based approximation method to simplify the characterization of angle of arrival uncertainties caused by UAV's fluctuation. Xiong \textit{et al.} \cite{10433063} developed a novel channel model for ARIS-assisted mmWave networks, explicitly accounting for UAV's tilt instability. Utilizing the refined model, they formulated a closed-form expression to characterize the signal-to-noise ratio under UAV's tilt. Furthermore, Xu \textit{et al.} \cite{10713302} proposed considering UAV's tilt to be an optimization variable to enhance the ergodic sum-rate in ARIS-assisted systems. By jointly optimizing the ARIS rotation in both elevation and azimuth angular dimensions, they formulated a dual-angle optimization problem and derived closed-form solutions. Despite these works demonstrating the impact of RIS orientation and UAV tilt on communication performance, few studies have integrated UAV's tilt into ARIS optimization.
	\subsection{Motivations and Contributions}
	Existing work predominantly neglects the critical impacts of altitude variations during ARIS flight and overlooks orientation-dependent performance degradation in communication systems. However, in practical scenarios, a quadrotor UAV inevitably experiences altitude variations due to inertial forces and acceleration, substantially constraining the achievable ARIS deployment gains. To address this challenge, we propose an Euler angles-based flight control paradigm integrated with quadrotor dynamics modeling. This framework enables simultaneous ARIS trajectory design and altitude optimization through control Euler angles, while maintaining optimal beamforming alignment via real-time phase shift adjustments. %The proposed methodology effectively bridges the  between aerial mobility constraints and electromagnetic wavefront manipulation requirements.
	
	Building upon the preceding discussion, the key contributions of this paper are outlined as follows:
	\begin{itemize}
		\item We investigate an ARIS-assisted communication system, where ARIS reflects signals from a BS to GUs. Given the impact of ARIS's altitude on performance gain, we propose an Euler angles-based ARIS control scheme for joint ARIS altitude and trajectory optimization. Therefore, we formulate an optimization problem to maximize the sum-rate by adjusting ARIS's altitude, trajectory, phase shifts, and BS beamforming, while ensuring compliance with constraints on BS transmission power, ARIS flight energy consumption, and flight safety.
		\item We transform the sum-rate maximization problem into a Markov decision process (MDP)-based model. Considering that the intractability of convex optimization-based methods and the limited exploration capabilities of conventional DRL algorithms in high-dimensional action space, a novel DRL framework based on the soft actor-critic with prioritized experience replay (SAC-PER) algorithm is proposed. The algorithm synergistically integrates maximum entropy reinforcement learning principles with stochastic policy optimization to enhance exploration efficiency while maintaining stable convergence.
		\item Numerical results demonstrate that the proposed Euler angles-based UAV control scheme effectively achieves joint altitude and trajectory optimization, exhibiting distinctly different trajectory compared to conventional horizontal ARIS baseline. Furthermore, the proposed SAC-PER outperforms benchmark methods in both learning efficiency and steady-state performance.
	\end{itemize}
	%\subsection{Organization and Notation}
	%The remaining sections of this paper are arranged as follows. In Section II, we describe the system model for ARIS-assisted communications. Section III formulates the sum-rate maximization problem by jointly optimizing the altitude, trajectory and phase shifts of the ARIS, along with the beamforming at the BS. Then, we reformulate the problem as an MDP framework. Section IV provides a detailed description of the proposed DRL algorithm. Numerical results are presented in Section V for evaluating the performance of the proposed scheme compared to benchmarks. Finally, conclusions are summarized in Section VI.
	
	\textit{Notation:} Scalars, vectors, and matrices are represented by italic letters, bold lowercase letters, and bold uppercase letters, respectively.
	 The collection of \(N \times M\) complex-valued matrices is symbolized as \(\mathbb{C}^{N \times M}\). For any complex-valued vector \(\mathbf{a}\), \(\|\mathbf{a}\|\), \(\mathbf{a}^T\), and \(\mathbf{a}^H\) indicate its Euclidean norm, transpose, and conjugate transpose, respectively. The expectation operator is written as \(\mathbb{E}[\cdot]\), and \(\mathrm{diag}(\mathbf{a})\) represents a diagonal matrix whose main diagonal entries are elements of \(\mathbf{a}\).
	
		\begin{table}[t]
		%\label{tab:parameters}
		\renewcommand{\arraystretch}{1.2} % Adjust row spacing
		\centering
		\caption{\scriptsize LIST OF VARIABLES}
		%\color{blue}
		\begin{tabular}{@{}c@{\hspace{0.01\textwidth}}p{0.37\textwidth}@{}}
			\toprule
			Variable & Description \\
			\midrule
			$K$ & $\text{The number of GUs}$ \\
			$N/{\bar N}$ & $\text{The number of ARIS/sub-surface elements}$ \\
			%$\bar N$ & $\text{The number of ARIS sub-surfaces}$ \\
			$M$ & $\text{The number of BS's antennas}$ \\
			$\mathbf{w}_k$ & $\text{The transmission beamforming at the BS for GU }k$ \\
			$L$ & $\text{Frame size }(\text{meter})$ \\
			$I_0$ & $\text{No-load current }(\text{A})$ \\
			$U_0$ & $\text{No-load voltage }(\text{V})$ \\
			$R_0$ & $\text{Motor resistance }(\Omega)$ \\
			$K_v$ & $\text{Nominal no-load motor constant  }(\text{rpm}/\text{V})$ \\
			$K_E$ & $\text{Back-electromotive force constant }$ ${K_E} \buildrel \Delta \over = \frac{{{U_0} - {I_0}{R_0}}}{{{K_v}{U_0}}}$ \\
			$K_T$ & $\text{Torque constant }$ ${K_T} \buildrel \Delta \over = 9.55{K_E}$ \\
			$P_{\rm BS}^{\rm max}$ & $\text{The maximum transmission power at the BS }(\text{W})$ \\
			$T$ & $\text{The duration of flight } (\text{s})$\\
			$L$ & $\text{The number of time slots}$\\
			$\delta$ & $\text{The length of each time slot } (\text{s})$\\
			$v_x/v_y/v_z$ & $\text{The speed of the ARIS on x-/y-/z-axis }(\text{m}/\text{s})$\\
			$a_x/a_y/a_z$ & $\text{The acceleration of the ARIS  on x-/y-/z-axis }(\text{m}/\text{s}^2)$\\
			%$F_{\rm tot}$ & $\text{The total lift of the aerial RIS }(\text{N})$\\
			$C_t$ & $\text{Thrust coefficient } (\text{N}/(\text{rad}/\text{s})^2)$\\
			$C_m$ & $\text{Torque coefficient } (\text{N}\cdot\text{m}/(\text{rad}/\text{s})^2)$\\
			$C_{dx}/C_{dy}/C_{dz}$ & $\text{Drag coefficient of x-/y-/z-axis } (\text{N}/(\text{m}/\text{s})^2)$\\
			$\omega_i$ & $\text{Speed of motor }i\text{ } (\text{rad}/\text{s})$\\
			$\phi/\theta/\psi$ & $\text{Roll/pitch/yaw angle } (\text{rad})$\\
			$\phi_{\rm max}/\theta_{\rm max}$ & $\text{Safety margin for } \phi/\theta \text{ } (\text{rad})$\\
			$\tilde\phi_{\rm max}/\tilde\theta_{\rm max}/\tilde\psi_{\rm max}$ & $\text{Safety variation for } \phi/\theta/\psi \text{ } (\text{rad})$\\
			$m$ & $\text{Aircraft mass } (\text{kg})$\\
			$g$ & $\text{The acceleration of gravity } (\text{m}/\text{s}^2)$\\
			$\alpha_k^{\rm RIS}/\alpha_{\rm BS}^{\rm RIS}$ & $\text{The azimuth from GU }k/ \text{BS to the ARIS} \text{ } (\text{rad})$\\
			$\beta_k^{\rm RIS}/\beta_{\rm BS}^{\rm RIS}$ & $\text{The elevation from GU }k/ \text{BS to the ARIS} \text{ } (\text{rad})$\\
			$K_1/K_{2}$ & $\text{The Rician factors}$\\
			%${\bf h}_{k}$ & $\text{The channel from the ARIS to GU $k$}$\\
			%$\bar{\bf h}_{k}/\tilde{\bf h}_{k}$ & $\text{The LoS/NLoS component of ${\bf h}_{k}$}$\\
			%${\bf{H}}$ & $\text{The channel from the BS to ARIS}$\\
			%$\bar{\bf H}/\tilde{\bf H}$ & $\text{The LoS/NLoS component of } {\bf H}$\\
			$d_{{\rm R},{\rm B}}/d_{{\rm R},k}$ & $\text{The distance between GU $k$/BS and the ARIS }(\text{m})$\\
			$\rho_0$ & $\text{The pass-loss factor at a reference distance (dBm)}$\\
			$\alpha_1/\alpha_2$ & $\text{The pass-loss exponents}$\\
			$H$ & $\text{The altitude of ARIS }(\text{m})$\\
			$D_{m}$ & $\text{The maximum directivity of the ARIS}$\\
			$G_k/G_{\rm B}$ & $\text{The reception/transmission gain}$\\
			$R_k$ & $\text{The achievable communication rate of GU }k$\\
			\bottomrule
		\end{tabular}
		\label{tab:notation}
	\end{table}

	\section{System Model and Problem Formulation}\label{s:sys}
	\begin{figure}[t]
		\centering
		\includegraphics[width=\columnwidth]{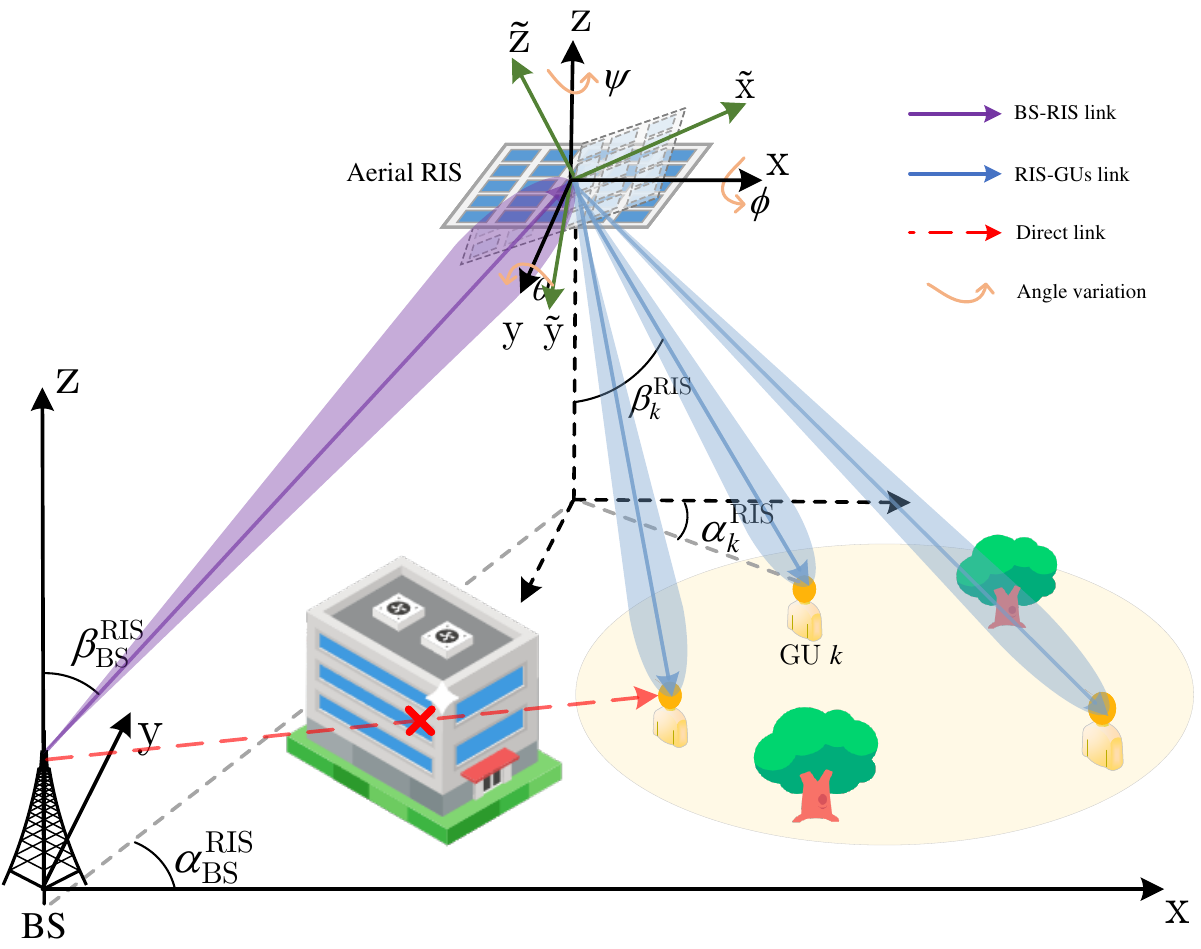}
		\captionsetup{justification=raggedright, singlelinecheck=false} % 设置标题靠左
		\caption{The system model of ARIS-assisted communication system with UAV altitude control.}
		\label{fig:system}
	\end{figure}
	In this section, we begin by introducing the ARIS-assisted communication system, where a BS with multiple antennas provides service to multiple single-antenna GUs with the ARIS.  Next, we present an Euler angles-based ARIS flight control framework and derive its associated flight energy consumption model. Building on these foundation, we analyze the practical ARIS channel gain and establish the signal transmission model.
	
	\subsection{Scenario Description}
	Considering an ARIS-assisted wireless communication system in which a BS equipped with $M$ antennas provides service to $K$ $(K \le M)$ single-antenna GUs. The set of GUs is denoted by $\mathcal{K}=\{ 1,\dots,k,\dots,K\}$. As depicted in Fig. 1, the potential obstacles may cause the direct links between the BS and GUs to be unreliable or even blocked. In response, an ARIS  composed of $N$ elements is introduced, denoted by $\mathcal{N}=\{ 1,\dots,n,\dots,N\}$, mounted on the UAV to establish high-quality communication links. Specially, the RIS is fixed beneath the UAV and tilting in accordance with the UAV's altitude. Let $T$ represent the flight duration of the UAV. For tractability, we partition $T$ into $L$ equal and non-overlapping time slots, each with length $\delta=T/L$. The set of time slots is represented by ${\cal L} = \left\{ {1, \ldots ,l, \ldots, L} \right\}$. The ARIS flies at a fixed altitude $H$ while continuously adjusting its Euler angles to achieve altitude and trajectory control. In each time slot, the position of the UAV is defined as $\mathbf{q}[l]=\left( {x[l],y[l],H} \right)$, the velocity is denoted by $\mathbf{v}[l]=\left( {v_x[l],v_y[l],0} \right)$, and the acceleration is $\mathbf{a}[l]=\left( {a_x[l],a_y[l],0} \right)$. Considering the practical scenario, the ARIS flight is subject to maximum speed and acceleration constraints as follows:
	\begin{equation}
		\left| \mathbf{v}[l] \right| \le {v_{\max }}, l \in \cal L,
	\end{equation}
	\begin{equation}
		\left| \mathbf{a}[l] \right| \le {a_{\max }}, l \in \cal L.
	\end{equation}
	
	Although introducing ARIS can significantly improve the communication quality, the BS-ARIS-GU links suffer from substantial path loss due to multiplicative fading, requiring a large number of ARIS elements to compensate. However, a large number of ARIS elements may cause excessive channel state information acquisition and ARIS design complexity. To solve this, the approach in \cite{UAV-STAR-RIS} is adopted to partition the $N$ elements into $\bar N$ sub-surfaces. Each sub-surface, indexed by the set $\mathcal{\bar N}=\{ 1,\dots,\bar n,\dots,\bar N\}$, consists of $\tilde{N}=N/\bar N$ (assumed to be an integer) adjacent elements sharing the same phase shift, thereby decreasing the overall implementation complexity. Specifically, for the $\bar n$-th sub-surface at time slot $l$, the reflection coefficient is given by $\theta_{\bar n}[l]=e^{j\varphi_{\bar n}[l]}$, where $\varphi_{\bar n}[l] \in \left[ {0,2\pi } \right)$ denotes the phase shift of this sub-surface. Therefore, the diagonal reflection coefficient matrix can be expressed as $\mathbf{\Theta}  = {\rm{diag}}\left( {\bm{\theta}[l] \otimes {\bm{1}_{\tilde N \times 1}}} \right) \in {\mathbb{C}^{N \times N}}$, where $\bm{\theta}[l]=\{ {\theta}_1[l],\dots,{\theta}_{\bar n}[l],\dots,{\theta}_{\bar N}[l]\}$, where $\otimes$ denotes the Kronecker product.
	%following the approach in \cite{UAV-STAR-RIS}, the $N$ elements are divided into $\bar N$ sub-surfaces, denoted by $\mathcal{\bar N}=\{ 1,\dots,\bar n,\dots,\bar N\}$, each consisting of $\tilde{N}=N/\bar N$ (assumed to be an integer) adjacent elements that share a common phase shift for reducing the overall implementation complexity. Specifically, the reflection coefficient of the $\bar n$-th sub-surface at time slot $t$ is denoted by $\theta_{\bar n}[t]=e^{j\varphi_{\bar n}[t]}$, where $\varphi_{\bar n}[t] \in \left[ {0,2\pi } \right)$ denotes the phase shift of the ${\bar n}$-th sub-surface. Therefore, the diagonal reflection coefficient matrix is given by $\mathbf{\Theta}  = {\rm{diag}}\left( {\bm{\theta}[t] \otimes {\bm{1}_{\tilde N \times 1}}} \right) \in {\mathbb{C}^{N \times N}}$, where $\bm{\theta}[t]=\{ {\theta}_1[t],\dots,{\theta}_{\bar n}[t],\dots,{\theta}_{\bar N}[t]\}$, $\otimes$ denotes the Kronecker product.
	\subsection{Dynamic Model of ARIS}
	%\begin{figure}[t]
	%	\centering
	%	\includegraphics[width=0.9\columnwidth]{figure/UAV_RIS_model.pdf}
	%	\captionsetup{justification=raggedright, singlelinecheck=false} % 设置标题靠左
	%	\caption{The model of ARIS.}
	%	\label{fig:UAV_RIS}
	%\end{figure}
	In this paper, we model the ARIS as a rigid body, with its Euler angles at time slot $l$ represented by the set $\Phi [l] = {\rm{\{ }}\phi [l],\theta [l],\psi [l]{\rm{\} }}$, where $\phi [l]$, $\theta [l]$, and $\psi [l]$ represent the roll, pitch, and yaw angles, respectively. The flight dynamics of the ARIS are powered by the continuous rotation of its four rotors. 
	%The rotors positioned opposite each other rotate in the same direction, whereas those next to each other rotate in opposite directions. 
	By adjusting the angular velocities of rotors, denoted by $\omega_i > 0, i \in \left \{1,2,3,4\right \}$ (only considering the magnitude of angular velocities), both trajectory and altitude control of ARIS can be achieved. According to \cite{9652043}, the thrust at time instant for each rotor is given by
	\begin{equation}
		F_i[l]=C_t\omega_i^2[l], i \in \left \{1,2,3,4\right \},
	\end{equation}
	where $C_t$ is the constant thrust coefficient.
	% 本文考虑的四旋翼无人机飞行动力来源于四个旋翼的不停旋转，如图2所示，相对的旋翼旋转方向相同，相邻的旋翼旋转方向相反，通过调节各旋翼的转速可以实现无人机的飞行以及姿态控制。在时间t上，记四个旋翼转速为\omega_i>0,i in {1,2,3,4} (只考虑转速大小，不考虑方向),依据文献[], 各旋翼升力
				
	% the total rotor angular velocity $\Omega$ and
	The dynamic model governing the ARIS flight control is described by
	\begin{equation}
		\left\{ {\begin{array}{*{20}{ll}}
				\displaystyle{m{a_x}[l] = {F_{{\rm{tot}}}}[l]\left( {\sin \psi [l]\sin \phi [l] + \sin \theta [l]\cos  [l]\cos \psi \cos\phi [l]} \right)}\\
				\displaystyle{\quad \quad \;\;\;\,\, - {C_{dx}}{v_x}[l]\left| {{v_x}[l]} \right|,}\\
				\displaystyle{m{a_y}[l] = {F_{{\rm{tot}}}}[l]\left( {\sin \theta [l]\sin \psi [l]\cos \phi [l] - \sin \phi [l]\cos \psi [l]} \right)}\\
				\displaystyle{\quad \quad \;\;\;\,\, - {C_{dy}}{v_y}[l]\left| {{v_y}[l]} \right|,}\\
				\displaystyle{m{a_z}[l] = {F_{{\rm{tot}}}}[l]\cos \phi [l]\cos \theta [l] - mg - {C_{dz}}{v_z}[l]\left| {{v_z}[l]} \right|,}
		\end{array}} \right.
	\end{equation}
	where the total thrust is calculated by
	\begin{equation}
		\displaystyle{F_{{\rm{tot}}}[l] = C_t \left ( \omega _1^2[l] + \omega _2^2[l] + \omega _3^2[l] + \omega _4^2[l]\right )}.
	\end{equation}
	
	As we consider the ARIS flight at a fixed altitude $H$, which implies that $v_z=0$ and $a_z=0$, the total thrust $F_{\rm tot}$ can be calculated by
	\begin{equation}
		{F_{{\rm{tot}}}}[l] = \frac{{mg}}{{\cos \phi [l]\cos \theta [l]}} .
	\end{equation}
	Consequently, given the ARIS's Euler angles, the accelerations along the $x$- and $y$-axes are given by
	\begin{equation}
		\scalebox{0.95}{$
		\displaystyle{a_x}[l] = \frac{{g\tan \phi [l]\sin \psi [l]}}{{\cos \theta [l]}}
			- g\tan \theta [l]\cos \psi [l] - \frac{{{C_{dx}}{v_x}[l]\left| {{v_x}[l]} \right|}}{m},
		$}
	\end{equation}
	\begin{equation}
		\scalebox{0.95}{$
		\displaystyle {a_y}[l] = g\tan \theta [l]\sin \psi [l] - \frac{{g\tan \phi [l]\cos \psi [l]}}{{\cos \theta [l]}} - \frac{{{C_{dy}}{v_y}[l]\left| {{v_y}[l]} \right|}}{m}.
		$}
	\end{equation}
	Therefore, both the ARIS's altitude and trajectory control can be realized.
					
	\subsection{Energy Consumption Model}
	Assuming uniform angular velocities for all rotors,  the angular velocity of each rotor can be obtained according to (5) and (6), given by
	\begin{equation}
		{\omega _i}[l] = \sqrt {\frac{{mg}}{{4{C_t}\cos \phi[l] \cos \theta[l] }}} ,i \in \left\{ {1,2,3,4} \right\}.
	\end{equation}		
	For each rotor, the corresponding current and voltage at each time slot are calculated by \cite{9652043}
	\begin{equation}
		I_i[l]=\frac{C_m}{K_T}\omega_i^2[l]+I_0,
	\end{equation}
	\begin{equation}
		U_i[l]=K_EN_i[l]+I_i[l]R_0.
	\end{equation}
	Therefore, the energy consumption of each motor can be obtained by
	\begin{equation}
		\begin{aligned}
			P_{i}[l]& =U_i[l]I_i[l] \\
			&=c_4\omega_i^4[l]+c_3\omega_i^3[l]+c_2\omega_i^2[l]+c_1\omega_i[l]+c_0,
		\end{aligned}
	\end{equation}
	where
	$c_0=I_0^2R_0$, $c_1=30K_EI_0/\pi$,  $c_2=2C_mR_0I_0/K_T$, $c_3=30C_mK_E/(\pi K_T)$, and $c_4=C_m^2R_0/K_T^2$.
	
	Combining equations (9) and (12), the flight energy consumption of the ARIS during time slot $l$ is given by 
	\begin{equation}
		\scalebox{1}{$
		\begin{array}{l}
			\displaystyle{P^{{\rm{fly}}}}[l] = \frac{{{c_4}}}{4}{\left( {\frac{{mg}}{{{C_t}\cos \phi[l] \cos \theta[l] }}} \right)^2} \\
			\displaystyle + \frac{{{c_3}}}{2}{\left( {\frac{{mg}}{{{C_t}\cos \phi[l] \cos \theta[l] }}} \right)^{\frac{3}{2}}} + \frac{{{c_2}mg}}{{{C_t}\cos \phi[l] \cos \theta[l] }} \\
			\displaystyle+ 2{c_1}{\left( {\frac{{mg}}{{{C_t}\cos \phi[l] \cos \theta[l] }}} \right)^{\frac{1}{2}}} + 4{c_0}.
		\end{array}
		$}
	\end{equation}
	Therefore, the sum energy consumption for ARIS can be calculated by ${E^{\rm {fly}}} = \sum\nolimits_{l = 1}^L {{P^{\rm {fly}}}[l]\delta }$.
	\subsection{Corresponding Angle Calculation}
	\begin{figure}[t]
		\centering
		\includegraphics[width=\columnwidth]{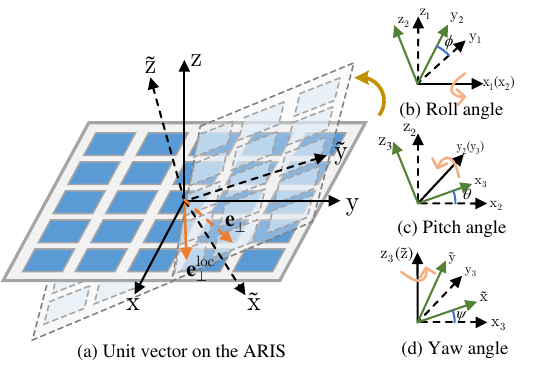}
		\captionsetup{justification=raggedright, singlelinecheck=false} % 设置标题靠左
		\caption{The altitude variation and definition of ARIS Euler angles.} %The axes in $\rm{x}_1\text{-}\rm{y}_1\text{-}\rm{z}_1$ are parallel to those in $\rm{x}\text{-}\rm{y}\text{-}\rm{z}$ of Fig. 1, the $\rm{{\rm{\tilde x}}}\text{-}\rm{\tilde y}\text{-}\rm{\tilde z}$ is defined on ARIS plane.}  
		\label{fig:fluctuation}
	\end{figure}
	%在UAV-RIS的local coordinate system(LCS)，the unit 
	%In the local coordinate system (LCS) ${\rm{\tilde x \text{-} \tilde y \text{-} \tilde z}}$ on the ARIS plane, as shown in Fig. 2(a), the unit vector pointing along the negative $\tilde z$-axis, also named the unit normal vector of the ARIS, is given by
	
	As shown in Fig. 2(a), the unit normal vector of the ARIS plane, aligned with the negative $\tilde z$-axis in the local coordinate system (LCS) ${\rm{\tilde x \text{-} \tilde y \text{-} \tilde z}}$, is defined as
	\begin{equation}
		e_ \bot ^{{\rm{loc}}} = {\left[ {\begin{array}{*{20}{c}}
					0&0&{ - 1}
			\end{array}} \right]^T}.
	\end{equation}
	%In this work, we focus on investigating the impact of ARIS flight attitude on its performance gain, where Euler angles are adopt. 
	Since the different coordinate frames are defined, the relationship between them, namely the coordinate transformation between global coordinate system and LCS, must be established. Firstly, the origin should be translated from $(0,0,0)$ to point $\left(x[l],y[l],H\right)$. Subsequently, the system undergoes sequential rotations: roll angle around ${\rm x}_1$-axis, pitch angle around ${\rm y}_2$-axis, and yaw angle around ${\rm z}_3$-axis, as shown in Fig. 2(b)-(d). Consequently, the transformation can be accomplished by multiplying the relevant rotation matrices, given by
	\begin{equation}
		\mathbf{R}_x\left( \theta [l] \right) = \left[ \begin{array}{ccc}
			\cos \theta [l] & 0 & \sin \theta [l] \\
			0 & 1 & 0 \\
			- \sin \theta [l] & 0 & \cos \theta [l]
		\end{array} \right],
	\end{equation}
	\begin{equation}
		\mathbf{R}_y\left( \phi [l] \right) = \left[ \begin{array}{ccc}
			1 & 0 & 0 \\
			0 & \cos \phi [l] & - \sin \phi [l] \\
			0 & \sin \phi [l] & \cos \phi [l]
		\end{array} \right],
	\end{equation}
	\begin{equation}
		\mathbf{R}_z\left( \psi [l] \right) = \left[ \begin{array}{ccc}
			\cos \psi [l] & - \sin \psi [l] & 0 \\
			\sin \psi [l] & \cos \psi [l] & 0 \\
			0 & 0 & 1
		\end{array} \right].
	\end{equation}
	
	\begin{figure*}
		\begin{equation}     
			%\begin{array}{l}
			%	\displaystyle {\bf{R}}[t] = {\bf{R}}_z\left( \psi [t] \right){\bf{R}}_y\left( \theta [t] \right){\bf{R}}_x\left( \phi [t] \right)\\
			%	\quad \quad  = \left[ \begin{array}{ccc}
			%		\cos \theta [t]\cos \psi [t] & \cos \psi [t]\sin \theta [t]\sin \phi [t] - \sin \psi [t]\cos \phi [t] & \cos \psi [t]\sin \theta [t]\cos \phi [t] + \sin \psi [t]\sin \phi [t] \\[7pt]
			%		\cos \theta [t]\sin \psi [t] & \sin \psi [t]\sin \theta [t]\sin \phi [t] + \cos \psi [t]\cos \phi [t] & \sin \psi [t]\sin \theta [t]\cos \phi [t] - \cos \psi [t]\sin \phi [t] \\[7pt]
			%		- \sin \theta [t] & \sin \phi [t]\cos \theta [t] & \cos \phi [t]\cos \theta [t]
			%	\end{array} \right].
			%\end{array}
			\begin{array}{*{20}{l}}
				\displaystyle{{\bf{R}}[l] = {{\bf{R}}_z}\left( {\psi [l]} \right){{\bf{R}}_y}\left( {\theta [l]} \right){{\bf{R}}_x}\left( {\phi [l]} \right)}\\
				{\quad \quad  = \left[ {\begin{array}{*{20}{ccc}}
							{\cos \psi [l]\cos \theta [l]}&{\cos \psi [l]\sin \phi [l]\sin \theta [l] - \sin \psi [l]\cos \phi [l]}&{\cos \psi [l]\cos \phi [l]\sin \theta [l] + \sin \psi [l]\sin \phi [l]}\\[5pt]
							{\sin \psi [l]\cos \theta [l]}&{\sin \psi [l]\sin \phi [l]\sin \theta [l] + \cos \psi [l]\cos \phi [l]}&{\sin \psi [l]\cos \phi [l]\sin \theta [l] - \cos \psi [l]\sin \phi [l]}\\[5pt]
							{ - \sin \theta [l]}&{\sin \phi [l]\cos \theta [l]}&{\cos \phi [l]\cos \theta [l]}
					\end{array}} \right].}
			\end{array}
		\end{equation}
		\end{figure*}
	
	The translation matrix could be obtained by multiplying these matrices, as shown in equation (18). Specifically, the unit normal vector ${\bf e}_\bot^{\rm loc}$ would be translated to 
	\begin{equation}	 
		\begin{array}{l}
			{{\bf e}_ \bot }[l] = {\bf{R}}[l]{\bf e}_ \bot ^{{\rm{loc}}}\\
			= \left[ {\begin{array}{*{20}{c}}
					{-\cos \psi [l]\cos \phi [l]\sin \theta [l] - \sin \psi [l]\sin \theta [l]}\\
					{-\sin \psi [l]\cos \phi [l]\sin \theta [l] + \cos \psi [l]\sin \theta [l]}\\
					{-\cos \phi [l]\cos \theta [l]}
			\end{array}} \right].
		\end{array}
	\end{equation}
	
	%在GCS下，GU k和BS到RIS的单位方向向量为
	The unit direction vectors of incident (between the BS and ARIS) and reflected signals (between the ARIS and GU $k$) are given by
	\begin{equation}	 
		{\bf e}_{k/{\rm BS}}^{\rm RIS}[l] = \left[ {\begin{array}{*{20}{c}}
				{\cos \beta _{k/{\rm BS}}^{\rm RIS}[l]\cos \alpha _{k/{\rm BS}}^{\rm RIS}[l]}\\[5pt]
				{\cos \beta _{k/{\rm BS}}^{\rm RIS}[l]\sin \alpha _{k/{\rm BS}}^{\rm RIS}[l]}\\[5pt]
				{\sin \beta _{k/{\rm BS}}^{\rm RIS}[l]}
		\end{array}} \right],
	\end{equation}
	where $\alpha _{k/{\rm BS}}^{\rm RIS}$ and $\beta_{k/{\rm BS}}^{\rm RIS}$ denote the azimuth and elevation angles from GU $k$ and the BS to the ARIS, respectively. Therefore, the angle between the incident/reflected signal and the normal vector of the ARIS plane can be obtained by
	% 从而，入射/出射信号与RIS面法向量夹角可根据下式进行计算
	\begin{equation}	 
		\begin{array}{l}
			\displaystyle \cos \gamma _{k/{\rm{BS}}}^{{\rm{RIS}}}[l] = \frac{-\boldsymbol{\mathit{\bf e}}_ \bot ^{{T}}[l]{\mkern 1mu} \boldsymbol{\mathit{\bf e}}_{k/{\rm{BS}}}^{{\rm{RIS}}}[l]}{ \left\| -\boldsymbol{\mathit{\bf e}}_ \bot [l] \right\| \, \left\| \boldsymbol{\mathit{\bf e}}_{k/{\rm{BS}}}^{{\rm{RIS}}}[l] \right\|} \\[10pt]
			= \cos \phi [l]\sin \theta [l]  \cos \beta _{k/{\rm BS}}^{\rm RIS}[l] \cos \left( \alpha _{k/{\rm BS}}^{\rm RIS}[l] - \psi [l] \right) \\[10pt]
			+ \cos \beta _{k/{\rm BS}}^{\rm RIS}[l] \sin \theta [l] \sin \left( \psi [l] - \alpha _{k/{\rm BS}}^{\rm RIS}[l] \right) \\[10pt]
			+ \sin \beta _{k/{\rm BS}}^{\rm RIS}[l] \cos \phi [l] \cos \theta [l].
		\end{array}
	\end{equation}
	This result highlights that the ARIS's altitude directly impact the incident and reflection angles, thereby altering ARIS gain and overall performance. 
	
	\subsection{Signal Model}
	For any time slot, the narrow-band quasi-static fading channels from the BS to ARIS, as well as from ARIS to GU $k$, denoted by ${\bf{H}}[l] \in {{\mathbb{C}}^{M \times N}}$ and $\mathbf{h}_{k}[l] \in {{\mathbb{C}}^{N \times 1}}$, are modeled as Rician fading channels, given by
	\begin{equation}
		\scalebox{0.95}{$
			\begin{aligned}
				\mathbf{H}[l] &= \sqrt{\frac{\rho_0}{{d_{\rm{R},\rm{B}}[l]}^{\alpha_1}}} 
				\left( 
				\sqrt{\frac{K_1}{1 + K_1}} \bar{\mathbf{H}}[l] + 
				\sqrt{\frac{1}{1 + K_1}} \tilde{\mathbf{H}}[l] 
				\right),
			\end{aligned}
			$}
	\end{equation}
	\begin{equation}
		\scalebox{0.95}{$
			\begin{aligned}
				{\bf{h}}_k[l]{\rm{ }} = \sqrt {\frac{{{\rho _0}}}{{{d_{{\rm{R}},k}}[l]^{{\alpha _2}}}}} \left( {\sqrt {\frac{{{K_{2}}}}{{1 + {K_{2}}}}} {\bf{\bar h}}_k[l] + \sqrt {\frac{1}{{1 + {K_{2}}}}} {\bf{\tilde h}}_k[l]} \right),
		\end{aligned}
		$}
		\end{equation}
	where $\rho_0$ represents the path loss at the reference distance of 1 meter, $\alpha_1$ and $\alpha_2$ are the pass loss exponents, ${d_{{\rm R},{\rm B}}}[l] = \left\| {{\mathbf{q}[l] - {\mathbf{q}_{\rm{B}}}}} \right\|$ is the distance between the ARIS and BS, ${d_{{\rm R},k}}[l] = \left\| {{\mathbf{q}}[l] - {\mathbf{q}_k}} \right\|$ is the distance between GU $k$ and the ARIS, with ${\mathbf{q}_{\rm{B}}}$ and ${\mathbf{q}_k}$ denote the position of the BS and GU $k$, respectively.  $K_1$ and $K_{2}$ denote the Rician factors, $\tilde{\mathbf{H}}[l]$ and ${\bf{\tilde h}}_k[l]$ are complex Gaussian random variables with independently and identically distributed zero mean and unit variance, $\bar{\mathbf{H}}[l]$ and ${\bf{\bar h}}_k^{{\rm{R}}}[l]$ represent the LoS components.%LoS部分是否需要表示 
	
	%补充RIS增益的计算,并且补充RIS无法覆盖的场景
	%考虑到RIS的实际增益受信号入射角和出射角，借助【】，RIS的实际增益可以建模为
	Considering that the practical channel gain of ARIS is influenced by the angles of signal incidence and reflection, the actual gain of the ARIS can be modeled as follows \cite{10444937}:
	\begin{equation}
		\begin{array}{l}
			{{\boldsymbol{\xi}}_k} = {G_k}[l]{G_{\rm{B}}}[l]{{\bf{\Phi }}_m}[l]\\
			\quad {\kern 1pt}  \buildrel \Delta \over = D_{{m}}^2F\left( {\upsilon _{k,{\rm{R}}}^{{\rm{AOD}}}[l],\vartheta _{k,{\rm{R}}}^{{\rm{AOD}}}[l]} \right)F\left( {\upsilon _{{\rm{R,B}}}^{{\rm{AOA}}}[l],\vartheta _{{\rm{R,B}}}^{{\rm{AOA}}}[l]} \right){{\bf{\Phi }}}[l],
		\end{array}
	\end{equation}
	where $D_{m}$ represents the ARIS's maximum directivity, $G_{k}$ signifies the reception gain from the BS to ARIS, and $G_{\rm{B}}$ represents the transmission gain from the ARIS to GU $k$. Additionally, $F\left(\upsilon,\vartheta\right)$ indicates the normalized power radiation pattern of the ARIS, with $\upsilon$ and $\vartheta$ denoting the azimuth and elevation angles between GU $k$ (BS) and the ARIS, respectively. This can be modeled using an exponential-Lambertian radiation pattern parameterized by $z$, which is given by
	\begin{equation}
		F\left( \upsilon , \vartheta \right) = \left\{\!\!\!
		\begin{array}{ll}
			\cos^z\left( \vartheta \right), & \upsilon \in \left[ 0, 2\pi \right], \vartheta \in \left[ 0, \pi \right], \\
			0, & \text{otherwise}.
		\end{array}
		\right.
	\end{equation}
	Based on equations (21), (24), and (25), the ARIS's gain for GU $k$ can is given by
	\begin{equation}
	\scalebox{1}{$
		{{\boldsymbol{\xi}}_k} = \left\{ \!\!\!\displaystyle{\begin{array}{*{20}{ll}}
				{D_{{m}}^2{{\left| {\cos \gamma _{{\rm{BS}}}^{{\rm{RIS}}}[l]\cos \gamma _k^{{\rm{RIS}}}[l]} \right|}^z}{\bf{\Theta }}[l],}&{\cos \gamma _{\rm{BS}/k}^{{\rm{RIS}}}[l] > 0,}\\
				\displaystyle{{{\bf{0}}_{N \times N}},}&{\text{otherwise}.}
		\end{array}} \right.
		$}
	\end{equation}
	
	Therefore, the received signal of GU $k$ is expressed as
	\begin{equation}	 
		{y_k}[l] = {\mathbf{v}_k}[l]{{\mathbf{w}}_k}[l]{x_k}[l] + \sum\limits_{j \ne k}^K {{\mathbf{v}_k}[l]{\mathbf{w}}_j[l]{x_j}[l]}  + n_k,
	\end{equation}
	where ${\mathbf{v}_k}[l] = \mathbf{h}_{k}^{H}[l]{{\boldsymbol{\xi}}_k} [l]{\mathbf{H}}[l] + \mathbf{h}_{{\rm BS},k}^{H}$ denotes the concatenated channel from the BS to GU $k$, ${\mathbf{w}}_k[l] \in \mathbb{C}^{M \times 1}$ is the $k$-th column of $\mathbf{W}[l] \in \mathbb{C}^{M \times K}$, which represents the BS's beamforming matrix, $x_k[l]$ is the transmission signal to GU $k$, satisfying $\mathbb{E}\left[ {{{\left| {{x_k}[l]} \right|}^2}} \right] = 1$, and ${n_k} \sim \mathcal{CN}\left( {0,{\sigma ^2}} \right)$ represents the additive Gaussian noise. Therefore, the achievable rate of GU $k$ is given by
	%\begin{equation}	 
	%	{\mathbf{h}_k}(t) = \mathbf{h}_{{\rm{R}},k}^H(t)\mathbf{\Phi} (t){\mathbf{V}}(t) + \mathbf{h}_{{\rm BS},k}^H
	%\end{equation}
	\begin{equation}	 
		{R_k}[l] = {\log _2}\left( {1 + \frac{{{{\left| {\mathbf{v}_k}[l]{\mathbf{w}_k}[l] \right|}^2}}}{{\sum\nolimits_{j \ne k}^K {\left| {\mathbf{v}_k}[l]{\mathbf{w}_j}[l] \right|}^2 + {\sigma ^2}}}} \right),
	\end{equation}
	The total sum-rate of all GUs over all time slots is expressed as
	\begin{equation}	 
		{R_{{\rm{sum}}}} = \sum\limits_{l = 1}^L {\sum\limits_{k = 1}^K {{R_k}[l]} }.
	\end{equation}
	
	\section{Problem Formulation and Markov Decision Process Model}\label{pro:s}
	In this section, we develop a sum-rate maximization problem that jointly optimizes the ARIS's altitude, trajectory, phase shifts and BS beamforming. We then model this problem as an MDP framework.
	
	\subsection{Problem Formulation}
	As indicated in equation (28), the achievable rate of GU $k$ is determined by the ARIS's position, altitude, phase shifts, and the BS beamforming. To investigate the impact of ARIS on communications, our goal is to maximize the sum-rate during the ARIS's flight duration through the joint optimization of the ARIS's Euler angles $\Phi$, reflection coefficient matrix $\bf \Theta$, and the BS beamforming matrix $\bf W$. In particular, the optimization problem is formulated as
	\begin{subequations}
		\begin{align}
			&\quad \;\mathop {\max }\limits_{{{\Phi}} ,{\bf {W}},{\bf{\Theta}} } \;{R_{{\rm{sum}}}}\\
			%&{\rm{s}}{\rm{.t}}{\rm{.}}\;{R_k}(t) \ge R_k^{\min },\forall k \in \mathcal{K},\forall t \in \mathcal{T},\\
			&{\rm{s}}{\rm{.t}}{\rm{.}}\;{\rm{Tr}}\left( {{{\bf {W}}^H}[l]{\bf {W}}[l]} \right) \le P_{{\rm{BS}}}^{\max },\forall l \in \mathcal{L},\\
			&\quad \;\;{\varphi _{\bar n}}[l] \in \left[ {0,2\pi } \right),\forall {\bar n} \in {\bar{\cal N}},\forall l \in \mathcal{L},\\
			%&\quad \;\;\phi [t] \in \left[ { - {\phi _{\max }},{\phi _{\max }}} \right],\forall t \in \mathcal{T},\\
			%&\quad \;\;\theta [t] \in \left[ { - {\theta _{\max }},{\theta _{\max }}} \right],\forall t \in \mathcal{T},\\
			&\quad \;\;\Phi [l] \in \left[ {{\Phi _{\min }},{\Phi _{\max }}} \right],\forall l \in {\cal L},\\
			&\quad \;\;\max \left\{ {\left| {\Phi [l + 1] - \Phi [l]} \right| - \tilde \Phi_{\rm max} } \right\} \le 0, l \le L - 1,\\
			&\quad \;\;{E^{{\rm{fly}}}} \le E_{{\rm{max}}}^{{\rm{fly}}},\\
			&\quad \;\;\min \left\{ {{\bf{q}}[l] - {{\bf{q}}_l}} \right\} \ge 0,l \in \mathcal{L},\\
			&\quad \;\;\max \left\{ {{\bf{q}}[l] - {{\bf{q}}_r}} \right\} \le 0,l \in \mathcal{L},\\
			&\quad \;\;(1),(2).
		\end{align}
	\end{subequations}
	Constraint (30b) ensures that the transmission power of the BS should not exceed the maximal transmission power. Constraint (30c) defines the feasible range of the ARIS's phase shifts. Constraints (30d) are established for flight safety consideration where ${\Phi _{\min }} = \left\{ { - {\phi _{\max }}, - {\theta _{\max }},0} \right\}$, ${\Phi _{\max }} = \left\{ {{\phi _{\max }},{\theta _{\max }},2\pi } \right\}$, imposing restrictions on the ARIS's pitch and roll angles, respectively. Constraint (30e) specifies the allowable variation in Euler angles between consecutive time slots, where ${\tilde \Phi _{\max }} = \{ {\tilde \phi _{\max }},{\tilde \theta _{\max }},{\tilde \psi _{\max }}\}$. Constraint (30f) governs the UAV's flight energy consumption. Constraints (30g) and (30h) specify that the ARIS can only move within a given range, where ${{\bf{q}}_l}$ and ${{\bf{q}}_r}$ represent the two vertices of the rectangular region. Constraint (30i) imposes limitations on the ARIS's flight speed and acceleration. 
	
	Problem (30) presents significant challenges for the following reasons. Firstly, the ARIS's altitude is intricately coupled with its flight trajectory, and optimizing the ARIS's altitude inevitably impacts its trajectory. Secondly, the gain of the ARIS is contingent upon the angles of signal incidence and departure, while the variation in ARIS's altitude and position further exacerbate the computational complexity associated with calculating the actual gain and optimizing the ARIS's phase shifts. Lastly, in uncertain environments, accurate online decision-making heavily relies on exhaustive environmental sampling during offline training. However, due to the practical limitations on feasible sampling, ensuring worst-case performance and guaranteeing safe online deployment emerge as additional formidable challenges. These factors make problems difficult to solve using traditional convex-based methods. Therefore, we adopt the SAC-PER-based algorithm to tackle these challenges.

	\subsection{MDP Formulation}
	In implementing DRL, we begin by defining the MDP which serves as the core structure for addressing sequential decision-making in uncertain environments. An MDP is characterized by a five-tuple $\left\{ {\mathcal{S},\mathcal{A},\mathcal{P},\mathcal{R},\gamma} \right\}$, where $\mathcal{S}$ is the set of environment states, $\mathcal{A}$ denotes the set of actions, $\mathcal{P}$ signifies the state transition probabilities, $\mathcal{R}$ represents the reward function, and $\gamma$ indicates the discount factor. At each time slot, the agent observes current state $s_l \in \cal{S}$ and selects an action $a_l \in \cal{A}$ following its stochastic policy $\pi (a_l|s_l) = P\left[ {{A_l} = a_l|{S_l} = s_l} \right] \in \left[ {0,1} \right]$. After receiving the action $a_l$, the environment transitions to the state $s_{l+1}$ and feeds back the reward $r_l$. The specific definitions for the state, action, reward, and state transition in our formulated MDP are provided below.
	
	%the agent selects an action $a \in \cal{A}$ from state $s \in \cal{S}$ according to its policy $\pi$. Specifically, the policy gives the probability distribution over actions for the given state, represented as $\pi (a|s) = P\left[ {{A_t} = a|{S_t} = s} \right] \in \left[ {0,1} \right]$. After executing action $a$, the agent transitions to the next state and receives a reward $r$. The specific definitions for the state, action, reward, and state transition in our formulated MDP are provided below.
	\subsubsection{State}
	At time slot $l$, the state is denoted by ${s_l} = \left\{ {\Phi [l],\mathbf{q}[l],\mathbf{v}[l],{R_{\rm sum}}[l],E_{{\rm{res}}}^{{\rm{fly}}}} \right\}$, which include the following five components:
	\begin{itemize}
		\item \textbf{} $\Phi [l] = \left\{ {\phi [l],\theta [l],\psi [l]} \right\}$: The set of ARIS's Euler angles at time slot $l$, including the roll, pitch, and yaw angles, respectively; 
		\item \textbf{} $\mathbf{q}[l]$: The position of the ARIS at time slot $l$;
		\item \textbf{} $\mathbf{v}[l]$: The velocity of the ARIS at time slot $l$;
		\item \textbf{} ${R_{{\rm{cum}}}}[l] = \sum\nolimits_{i = 1}^{l - 1} {\sum\nolimits_{k = 1}^K {{R_k}[i]} } $: The sum-rate of all GUs from time slot 1 to $l-1$;
		\item \textbf{} $E_{{\rm{rem}}}^{{\rm{fly}}}[l] $: The remaining flight energy of the ARIS.
	\end{itemize}
	%计算状态空间大小
	
	\subsubsection{Action}
	The formulated MDP's action space consists of the ARIS's Euler angles, phase shifts of each sub-surface, and BS beamforming decision at each time slot. Given the above action space, determining the optimal policy poses critical challenges due to the following factors. Firstly, for flight safety considerations, the variation and maximum values of the ARIS's Euler angles in each time slot are subject to constraints (30e), (30f), and (30g). Directly using Euler angles as optimization variables makes it challenging to simultaneously satisfy both of these constraints. Additionally, the high-dimensional action space and environmental uncertainties render the MDP difficult to solve, as the transition probabilities are unknown, and the curse of dimensionality further complicates the optimization process. To address the above challenging issues, we treat the variation in Euler angles as optimization variables, denoted as $\tilde \Phi  = \left\{ {\tilde \phi ,\tilde \theta ,\tilde \psi } \right\}$. To satisfy constraint (30g), we impose bounds on their values, i.e. $\max \left\{ {\left| {\tilde \Phi } \right| - {{\tilde \Phi }_{\max }}} \right\} \le 0$. Furthermore, to meet constraints (30e) and (30f), after the agent selects an action, we adjust the action based on current Euler angles to ensure compliance with these constraints. 
	Additionally, to keep the action relatively small, a low-complexity method is proposed to design the BS beamforming matrix under the given ARIS's altitude, position, and phase shifts. The details of this approach are presented as follows.
	%我们将欧拉角的变化值作为优化变量，表示为$\tilde \Phi  = \left\{ {\tilde \phi ,\tilde \theta ,\tilde \psi } \right\}$，为了满足约束(31g)，对其取值范围进行约束，即$\left| {\tilde \alpha } \right| \le {\tilde \alpha _{\max }},\tilde \alpha  \in \tilde \Phi$，同时，为了满足约束(31e)和(31f),我们在智能体选择动作后，会根据当前欧拉角的值对动作进行修正以满足约束。
	%首先，处于安全考虑，aerial RIS的欧拉角各个时隙的变化量和最大值有约束，直接将欧拉角作为优化变量难以同时满足这两种约束。其次，Secondly, the high-dimensional action space and the environment uncertainties make the MDP quite a challenge to solve due to the unknown transition probabilities and the curse of dimensionality
	
	%\paragraph{Optimization for resource allocation and transmission power}
	
	Since the BS beamforming matrix is independent across different time slots, we omit the time slot $l$ in the beamforming matrix derivation for simplicity. At a particular time slot, once the ARIS's altitude, position and phase shifts are given, the BS beamforming optimization subproblem can be reformulated as
	\begin{subequations}
		\begin{align}
			&\;\;\;\;\;\mathop {\max }\limits_\mathbf{W} \;{R_{{\rm{sum}}}}\\
			&{\rm{s}}.{\rm{t}}.\;{\rm{Tr}}\left( {{\mathbf{W}^H}\mathbf{W}} \right) \le P_{{\rm{BS}}}^{\max }.
			%&\quad \;\;{\rm{Tr}}\left( {{\mathbf{W}^H}\mathbf{W}} \right) \le P_{{\rm{BS}}}^{\max }.
		\end{align}
	\end{subequations}
	To address the digital beamforming optimization problem (31), zero-forcing (ZF) precoding, a low-complexity strategy that can effectively eliminate multi-user interference while achieving the near-optimal performance, is employed. The received signal in equation (27) can be rewritten as $\bf{y} = \bf{V}\bf{W}\bf{x} + \bf{n}$, where we have ${\bf{y}} = \left[ {{y_1}, \ldots ,{y_K}} \right]^T$, ${\bf{x}} = \left[ {{x_1}, \ldots ,{x_K}} \right]^T$, $\bf{V}$ denotes a $K \times M$ matrix with the $k$-th row being ${\bf{v}}_k$, and $\bf{n}$ is the noise vector. The ZF beamforming matrix is calculated by
	\begin{equation}
		\mathbf{W} = {\mathbf{V}^H} \left( \mathbf{V} \mathbf{V}^H \right)^{-1} \mathbf{P}^{\frac{1}{2}} = \tilde{\mathbf{V}} \mathbf{P}^{\frac{1}{2}},
	\end{equation}
	where $\tilde{\mathbf{V}} = {\mathbf{V}^H}{\left( {\mathbf{V}{\mathbf{V}^H}} \right)^{ - 1}}$, and $\bf{P}$ is a diagonal matrix with the $k$-th diagonal element being $p_k$, calculated by
	\begin{equation}
		{p_k} = \frac{1}{{{\nu_k}}}\max \left\{ {\frac{1}{\mu } - {\nu_k}{\sigma ^2},0} \right\},
	\end{equation}
	where $\nu_k$ represent the $k$-th diagonal element of ${\widetilde {\bf{V}}^H}\widetilde {\bf{V}}$, and $\mu$ serves as a normalization factor chosen to ensure
	\begin{equation}
		\sum\limits_{k = 1}^K {\max \left\{ {\frac{1}{\mu } - {\nu_k}{\sigma ^2},0} \right\} = P_{{\rm{BS}}}^{\max }}.
	\end{equation}
	
	%为了求得最优拉格朗如乘子，我们设计了一种基于二分法和
	Considering the ARIS's altitude, some GUs may fall outside the service half-space of ARIS, leading to obstructed communication links between these GUs and the BS. This makes it challenging to determine the feasible bounds of the normalization factor $\mu$, causing prohibitively high computational complexity in solving for the optimal $\mu$ via the bisection method. To mitigate this issue, we introduce a service factor $\kappa_k$ prior to conducting the bisection method, given by
	\begin{equation}
		{\kappa _k}[l] = \left\{ {\begin{array}{*{20}{ll}}
				{D_m^2{{\left| {\cos \gamma _{{\rm{BS}}}^{{\rm{RIS}}}[l]\cos \gamma _k^{{\rm{RIS}}}[l]} \right|}^z},}&{\cos \gamma _{{\rm{BS/}}k}^{{\rm{RIS}}}[l] > 0,}\\
				{0,}&{{\rm{otherwise}}.}
		\end{array}} \right.
	\end{equation}
	When $\kappa_k[l] > \kappa_{\min}$, the ARIS effectively covers GU $k$ within its half-space. This condition is employed as a criterion when determining the feasible bound for the bisection method. The algorithm is summarized in Algorithm 1.
	%and $p_k^{\min } = {\sigma ^2}\left( {{2^{R_k^{\min }}} - 1} \right)$ is the minimum received power constraint of GU $k$.
	
	\begin{algorithm}[t]
		\caption{Water-Filling and Bisection-Based Algorithm for Solving (32)}
		\begin{algorithmic}[0]
			\STATE {\textbf{Input:}} ${\mathbf{h}}_{{\rm R},k}$, ${\mathbf{h}}_{{\rm BS},k}$, $\mathbf{H}$, ${{\boldsymbol{\xi}}_k}$, $\sigma ^2$, $ \kappa_{\min}=10^{-4}$
			\STATE {\textbf{1. Initialization:}} \\
			Calculate matrix $\tilde{\bf V}^H \tilde{\bf V}$ and obtain $\nu_k$ for each GU\\
			Initialize $\mu_{\rm max} = \mu_{\rm min} = \mu_{\rm init}$
			\STATE {\textbf{2. Finding upper and lower bounds for $\mu$:}} \\
			\FOR{$k \le K$}
			\STATE \textbf{if} \scalebox{1}{{$\nu_k \sigma^2 \le 1/{\mu _{\max }}$} and $\kappa_k>\kappa_{\rm min}$ \textbf{then} ${\mu _{\max }} = 1/{\nu_k}{\sigma ^2}$}
			\STATE \textbf{if} \scalebox{1}{{$\nu_k \sigma^2 > 1/{\mu _{\min }}$} and $\kappa_k>\kappa_{\rm min}$ \textbf{then} ${\mu _{\min }} = 1/{\nu_k}{\sigma ^2}$}
			\ENDFOR
			\STATE {\textbf{3. Finding the optimal $\mu$ based on bisection method:}} \\
			\STATE \textbf{repeat}\\
			\STATE \hspace{1em} Calculate the middle value ${\mu _{{\rm{mid}}}} = \left( {{\mu _{{\rm{max}}}} + {\mu _{{\rm{min}}}}} \right)/2$\\
			\STATE\hspace{1em} \textbf{if} \scalebox{1}{$\sum\nolimits_{k = 1}^K {\max \left\{ {\frac{1}{{{\mu _{{\rm{mid}}}}}} - {\nu_k}{\sigma ^2},0} \right\}}  > P_{{\rm{BS}}}^{\max } $} \\
			\STATE\hspace{2em} \textbf{then} \scalebox{1}{$\mu_{\min}=\mu_{\rm mid}$}\\
			\STATE\hspace{1em} \textbf{else if} $\sum\nolimits_{k = 1}^K {\max \left\{ {\frac{1}{{{\mu _{{\rm{mid}}}}}} - {\nu_k}{\sigma ^2},0} \right\}}  < P_{{\rm{BS}}}^{\max } $ \\
			\STATE\hspace{2em} \textbf{then} \scalebox{1}{$\mu_{\max}=\mu_{\rm mid}$}\\
			\STATE\hspace{1em} \textbf{else break}\\
			\STATE {\textbf{4. Obtaining the optimal beamforming based on (33)}}
		\end{algorithmic}
	\end{algorithm}
	
	From equations (32) and (33), the optimal BS beamforming matrix is derived under given ARIS's altitude, position, and phase shifts. Consequently, in our MDP formulation, only the ARIS's phase shifts and the variations of Euler angles need to be involved in the action space, while the optimal BS beamforming is determined based on equations (32) and (33) to facilitate state-value computation. Therefore, the action space consists of two components as follows: 
	\begin{itemize}
		\item \textbf{} $\tilde \Phi[l]  = \left\{ {\tilde \phi[l] ,\tilde \theta[l] ,\tilde \psi[l]} \right\}$: The variation of ARIS's Euler angles at time slot $l$;
		\item \textbf{} $\left\{ {{\varphi _1}[l], \ldots ,{\varphi _{\tilde n}}[l], \ldots ,{\varphi _{\tilde N}}[l]} \right\}$: The phase shifts of ARIS's sub-surfaces at time slot $l$.
	\end{itemize}

	\subsubsection{Reward}
	As stated in (30), the objective of optimizing ARIS's altitude, trajectory, phase shifts, and BS beamforming matrix is to maximize the sum-rate across all time slots.
	To align with this objective, the reward guiding the learning should incorporate all GUs's instantaneous sum-rate at each time slot, namely $\bar R[l] = \sum\nolimits_{k = 1}^K {{R_k}[l]}$. To address the flight range constraint, we introduce a penalty $P_1$ when the ARIS exits the designated rectangular region. Furthermore, to account for the energy consumption constraint during flight, we incorporate a penalty term $\omega E_{\text{res}}^{\text{fly}}$ when the ARIS's remaining flight energy becomes negative. Finally, to enforce the maximum speed and acceleration constraints during ARIS flight, we introduce penalty terms $P_3$ and $P_4$, respectively. Thus, the reward function is defined as follows:
	\begin{equation}
		{r_t} = 
		\left\{
		\!\!\!\! \begin{array}{l@{\hspace{10pt}}l}
			\bar{R}[l] - P_1, & \text{if} \; \min \left\{ \mathbf{q}[l] - \mathbf{q}_l \right\} < 0, \\[5pt]
			\bar{R}[l] - P_1, & \text{if} \; \max \left\{ \mathbf{q}[l] - \mathbf{q}_r \right\} > 0, \\[5pt]
			\bar{R}[l] + \omega E_{\text{res}}^{\text{fly}}, & \text{if} \; l < L \; \text{and} \; E_{\text{res}}^{\text{fly}} < 0, \\[5pt]
			\bar{R}[l] - P_2, & \text{if} \; {\bf v}[l] > {\bf v}_{\max}, \\[5pt]
			\bar{R}[l] - P_3, & \text{if} \; {\bf a}[l] > {\bf a}_{\max}.
		\end{array}
		\right.
	\end{equation}
	%其中$R( \cdot )$表示求矩阵的秩。
	Note that parameters $P_1$, $P_2$, $P_3$, and $\omega$ should be finely adjusted to enhance both the the expected accumulated reward and convergence performance.
	\subsubsection{State Transition}
	After the agent selects an action, the state is updated accordingly. Firstly, the ARIS's Euler angles are updated based on the determined variation, given by
	\begin{equation}
		\Phi [l+1] = \Phi [l] + \tilde \Phi [l].
	\end{equation}
	Next, the ARIS's acceleration during this time slot can be computed using equations (7) and (8), and the velocity is updated as
	\begin{equation}
		{\bf v}[l+1] = {\bf v}[l] + {\bf a}[l]\delta.
	\end{equation}
	Using the updated acceleration, the ARIS's position is updated by
	\begin{equation}
		{\bf{q}}[l + 1] = {\bf{q}}[l] + {\bf{v}}[l]\delta  + \frac{1}{2}{\bf{a}}[l]{\delta ^2}.
	\end{equation}
	Given the ARIS's altitude and position, the transmission rate for each user can be computed using equation (28), and the cumulative rate is updated by
	\begin{equation}
		{R_{{\rm{cum}}}}[l+1] = {R_{{\rm{cum}}}}[l] + \bar R[l].
	\end{equation}
	Finally, the ARIS's flight energy consumption at this time slot can be computed using equation (13), and the remaining flight energy is updated by
	\begin{equation}
		E_{{\rm{rem}}}^{{\rm{fly}}}[l+1] = E_{{\rm{rem}}}^{{\rm{fly}}}[l] - {P^{{\rm{fly}}}}[l]\delta.
	\end{equation}

	\subsection{SAC-Based Algorithm}
	\subsubsection{SAC framework}
	Although DRL has been highly anticipated for real‐world applications, its progress remains slow, largely due to limited sampling efficiency and unstable convergence \cite{UAV-STAR-RIS}. To address these issues, the SAC framework, grounded in the maximum entropy principle, was introduced to promote sample efficiency in training. Compared with conventional DRL methods, SAC provides multiple benefits, including multi-mode near-optimal policies, more efficient exploration, and faster training speed, particularly for challenging tasks. In standard DRL frameworks, the optimization objective is to maximize the expected cumulative rewards from the initial state.
	Let the policy $\pi$ induce a state-action trajectory distribution denoted by ${\rho _\pi }$. Thus, the agent's objective can be expressed as
	\begin{equation}
		\mathop{\max}\limits_\pi \; \sum_{l=1}^{L} \mathbb{E}_{(s_l, a_l) \sim \rho_\pi} \left[ \gamma^{l - 1} r(s_l, a_l) \right].
	\end{equation}
	The SAC framework incorporates an entropy term into the objective function to encourage exploration. Specifically, the objective is formulated as
	\begin{equation}
		\sum\limits_{l=1}^{L} \mathbb{E}_{(s_l, a_l) \sim \rho_\pi} \left[ \gamma^{l - 1} r(s_l, a_l) + \alpha \mathcal{H}(\pi(\cdot| s_l)) \right],
	\end{equation}
	where $\alpha\mathcal H(\pi(\cdot | s_l)) = - \mathbb{E}_{a \sim \pi(\cdot | s_l)} \log_2 \pi(a | s_l)$ denotes the entropy of policy distribution, with the temperature hyperparameter $\alpha$ regulates the weight of the entropy and reflects the degree of stochasticity in the optimal policy $\pi^{*}$. 
	
	The SAC framework is fundamentally based on the policy iteration algorithm, including two primary phases: policy evaluation and policy improvement. Within the evaluation phase, the action values for a given policy $\pi$ are assessed by the Bellman expectation function, given by $Q_\pi(s_l, a_l) = r(s_l, a_l) + \gamma \mathbb E_{s_{l+1} \sim \rho_{\pi}} \left[ v_\pi(s_{l+1}) \right]$. Compared to the traditional DRL algorithms, by involving the entropy, the state-value function of SAC is given by
	\begin{equation}
		{v_\pi }\left( {{s_l}} \right) = {\mathbb E_{{a_l} \sim \pi }}\left[ {{Q_\pi }\left( {{s_l},{a_l}} \right) - \alpha {{\log }_2}\left( {\pi \left( {{a_t}\left| {{s_l}} \right.} \right)} \right)} \right].
	\end{equation}
	
	Given that the state space in our proposed MDP is continuous, neural networks are employed to approximate the state values. Let $\omega$ represent the parameters of the Q-network. Then, its loss function is expressed as
	\begin{equation}
		\scalebox{1}{$
			{L_Q}\left( \omega  \right) = {\mathbb E_{\left( {{s_l},{a_l}} \right) \sim {\cal D}}}\left[ {\frac{1}{2}{{\left( {{Q_\omega }\left( {{s_l},{a_l}} \right) - {{\hat Q}}\left( {{s_l},{a_l}} \right)} \right)}^2}} \right],
			$}
	\end{equation} 
	where 
	\begin{equation}
		\scalebox{0.96}{$
			\begin{array}{l}
				\hat Q\left( s_l, a_l \right) = r\left( s_l, a_l \right) + \gamma \sum\limits_{{a_{l + 1}} \in A} {\displaystyle \pi \left( a_{l + 1} | s_{l + 1} \right)} \\
				\quad \quad \quad \quad\, \times \left[ Q_{\hat{\omega}}\left( s_{l + 1}, a_{l + 1} \right) - \alpha \log \left( \pi \left( a_{l + 1} | s_{l + 1} \right) \right) \right].
			\end{array}
			$}
	\end{equation}
	Here, $\cal D$ represents the replay buffer, $\hat \omega$ is the parameter of target Q-network, which is periodically copied from $\omega$.
	
	The policy improvement iteratively enhances the policy $\pi$ by leveraging real-time Q-values estimated from policy evaluation. The loss function for the network is given by
	\begin{equation}
		\scalebox{1}{$
			{L_\pi }\left( \varphi  \right) = {\mathbb{E}_{{s_l} \sim {\cal D}}}{\mathbb{E}_{{a_l} \sim {\pi_\varphi}}}\left[ \alpha {{\log}_2}\!\left( {\pi_\varphi}\left( {a_l\left| {s_l} \right.} \right) \right) - {Q_\omega }\left( {s_l},{a_l} \right) \right].
			$}
	\end{equation}
	
	\subsubsection{Temperature Auto-adjustment}
	SAC is highly sensitive to the temperature coefficient of entropy, as it controls the balance between reward and entropy, influencing the algorithm's ability to explore and exploit. In the early state of training, the temperature $\alpha$ should be increased to encourage better exploration. As the training progresses, a smaller $\alpha$ can allow agent to make more effective use of high-quality samples. In order to accomplish this, we leverage the recursive form of $ \mathbb{E}_{(s_l, a_l) \sim \rho_\pi} \left[ \gamma^{l-1} r(s_l, a_l) \right]$ and apply the strong duality principle. Consequently, the optimal dual variable $\alpha^{*}_l$ is given by
	\begin{equation}
		\alpha_l^* = \mathop {\arg \min }\limits_{\alpha_l} \mathbb{E}_{\alpha_l \sim \pi_l^*}\left[ - \alpha_l \log \left( \pi_l^* \left( a_l \mid s_l ; \alpha_l \right) \right) - \alpha_l \mathcal{H}_{\min} \right],
	\end{equation}
	where ${\pi _l^*\left( {{a_l}\left| {{s_l}} \right.;{\alpha _l}} \right)}$ represents the optimal policy under the temperature $\alpha_l$, $\mathcal{H}_{\min}$ denotes the minimum-entropy constraint. Therefore, dual gradient descent stands out as a viable approach, with the objective being 
	\begin{equation}
		{L}(\alpha) = \mathbb{E}_{a_l \sim \pi_l}\left[ -\alpha \log \left( \pi_l \left( A_l \mid S_l \right) \right) - \alpha \mathcal{H}_{\min} \right].
	\end{equation}
	
	\begin{algorithm}[t]
		\caption{Our proposed SAC-PER algorithm}
		\label{SAC}
		\begin{algorithmic}[1]
			\STATE{Initialize the environment.}
			\STATE{
				Initialize critic network parameters ${\omega _i}(i = 1,2)$ and actor network parameter $\varphi$.}
			\STATE{
				Set entropy level $\mathcal H_{\min }$, replay buffer ${\cal D} = \emptyset$, learning rate, temperature parameter $\alpha$, and discount factor $\gamma$, respectively.}
			\FOR{each episode}
			\FOR{each environment step}
			\STATE{Select action $a_l$ based on current policy.}
			\STATE{Take action $a_l$ and calculate the ARIS's altitude and position based on equations (37) and (39). Then, use equations (21) and (26) to compute the gain of ARIS. Finally, apply Algorithm 1 to obtain the optimal BS beamforming matrix.%执行动作a_t，并根据式(37)和(39)计算出ARIS的姿态与位置，接着利用式(22)和(27)计算出ARIS的实际增益，利用算法1求得最优基站波束赋形矩阵。
				\STATE{Transmit to the next state $s_{l+1}$, calculate the reward $r_l$ and then store transition tuple $\left\{ s_l, a_l, r_l, s_{l+1} \right\}$ in the $\cal D$.}
				\STATE \textbf{if} Sample size meets the requirement of $N_b$ \textbf{do}
				\STATE\hspace{1em} \textbf{for } $b \in {\cal{B}}_{\rm {batch}}$ \textbf{do}
				\STATE\hspace{2em} Sample $i$ with probability $P_i$.
				\STATE\hspace{2em} Calculate importance sampling by (53).
				\STATE\hspace{2em} Calculate TD-error $\delta_i$ by (51).
				\STATE\hspace{2em} Calculate priority $p_i$.
				\STATE\hspace{1em} \textbf{end for}
				\STATE \textbf{end if}
			}
			\ENDFOR
			\FOR{each gradient step}
			\STATE{Update critic networks $\omega_i$ by loss function (45):\\${\omega _i} \leftarrow {\omega _i} - \lambda {\nabla _{{\omega _i}}}{L_Q}\left( {{\omega _i}} \right),i \in \left\{ {1,2} \right\}$.}
			\STATE{Update the actor network $\varphi$ by loss function (47):\\$\varphi  \leftarrow \varphi  - \lambda {\nabla _\varphi }{L_\pi }\left( \varphi  \right)$.}
			\STATE{Update temperature $\alpha$ by solving (48):\\$\alpha  \leftarrow \alpha  - \lambda {\nabla _\alpha }L\left( \alpha  \right)$.}
			\STATE{Update target network parameter $\hat \omega_i$:\\${\hat \omega _i} \leftarrow \tau {\omega _i} + \left( {1 - \tau } \right){\hat \omega _i},i \in \{ 1,2\}$.}
			\ENDFOR
			\ENDFOR
		\end{algorithmic}
	\end{algorithm}
	\subsubsection{Prioritized Experience Replay (PER)}
	In contrast to traditional experience replay mechanisms, we employ PER to improve the training efficiency in DRL frameworks. Specifically, each transition is prioritized according to its temporal difference error (TD-error), which quantifies the discrepancy between the value predicted by the current model and the target value of the sample. Transitions with larger TD-error values are deemed more critical for model updates, as they indicate regions where the model’s predictions are less accurate. The implementation of a prioritized sampling mechanism, which selectively experience data based on estimated sample importance, enables more efficient neural network training by focusing computational resources on high-impact transitions. %This strategy optimizes the training process of neural networks by ensuring that critical samples are given precedence, thereby improving the overall efficiency and performance of the DRL framework.

	Taking DQN with PER as an example, the TD-error for each experience tuple is calculated based on the interpolation between the current and target $Q$ values, given by
	\begin{equation}
		{\delta _l} = r\left( {{s_l},{a_l}} \right) + \gamma {Q_{{\rm{target}}}}\left( {{s_{l + 1}},{a_{l + 1}}} \right) - Q\left( {{s_l},{a_l}} \right),
	\end{equation}
	where $Q_{{\rm{target}}}$ denote the target $Q$ network, and $Q$ is the current $Q$ network. As the SAC algorithm contains two $Q$-network, the TD-error is set as the mean absolute value of the TD-error for the two $Q$-network, which is expressed as
	\begin{equation}
		\left| {{\delta _l}} \right| = \frac{1}{2}\sum\limits_{i = 1}^2 {\left| {{Q_{{\omega _i}}}({s_l},{a_l}) - {Q_{{\rm{target}}}}({r_l},{s_{l + 1}})} \right|}.
	\end{equation}
	Therefore, the sampling probability for sample $i$ is given by
	\begin{equation}
		P(i) = \frac{{p_i^{{\beta _1}}}}{{\sum\nolimits_k {p_k^{{\beta _1}}} }},
	\end{equation}
	where $\beta_1$ is the distribution factor, and $p_i$ denotes the priority of sample $i$, calculated by ${p_i} = \left| {{\delta _i}} \right| + \varepsilon$, with $\varepsilon$ denoting a positive constant to prevent the priority $p_i$ from becoming zero. Since the prioritized replay alters sample's likelihood of being drawn, an importance sampling weight $w_i$ must be introduced to adjust the error updates, given by
	\begin{equation}
		{w_i} = {\left( {\frac{1}{N_D} \cdot \frac{1}{{P\left( i \right)}}} \right)^{{\beta _2}}},
	\end{equation}
	where $N_D$ denotes the capacity of the experience replay, 
	$\beta_2$ is a constant value for adjusting sampling weight \cite{9913936}, satisfying ${\beta _2} \in \left[ {0,1} \right]$. When $\beta_2$ is equal to 0, the importance sampling is not used, and when $\beta_2$ is equal to 1, the impact of PER on convergence is completely offset. Fig. 3 and Algorithm 2 illustrate the architecture and training process of proposed SAC-PER algorithm.
	%\subsection{Complexity Analysis}

	%SAC算法复杂度分析
	%\begin{figure}[t]
	%	\centering
	%	\includegraphics[width=\columnwidth]{figure/network.pdf}
	%	\captionsetup{justification=raggedright, singlelinecheck=false} % 设置标题靠左
	%	\caption{Architecture of SAC-based algorithm.}
	%	\label{fig:network}
	%\end{figure}
	\begin{figure*}[t]
		\centering
		\captionsetup{justification=raggedright, singlelinecheck=false}
		\includegraphics[width=1.75\columnwidth]{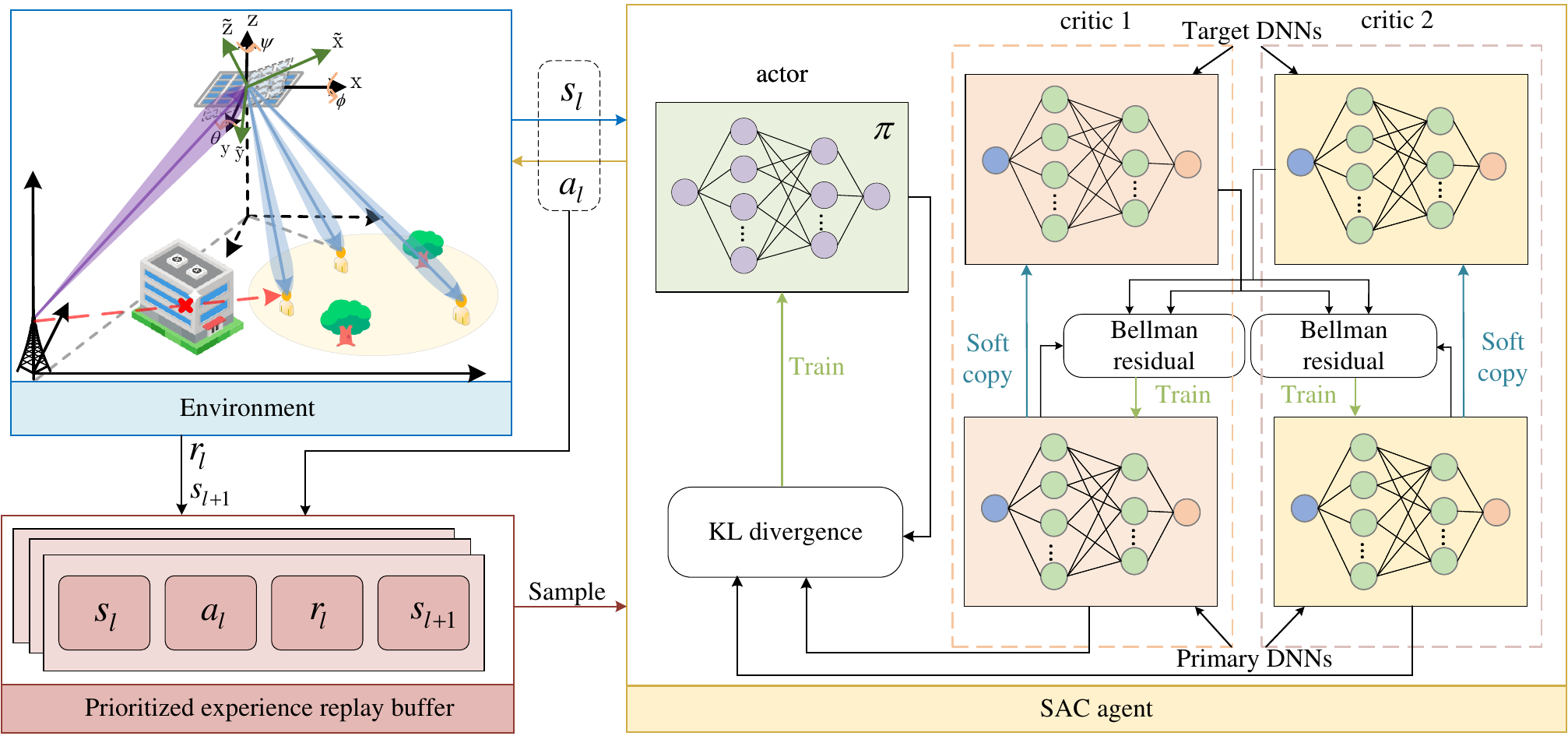}
		\caption{The SAC-PER framework.}
		\label{fig:network}
	\end{figure*}
	
	\section{Complexity Analysis}
	Within the proposed SAC-PER algorithm, the complexity mainly arises from training actor and critic networks. Specially, the training complexity arises from the forward and backward propagation performed in DNNs. Since the complexity of backward propagation is comparable to that of forward propagation, the time complexity of network training is ${\cal O}\left( {\sum\limits_{i = 0}^{I - 1} {{l_i}{l_{i + 1}}}  + \sum\limits_{j = 0}^{J - 1} {{{\hat l}_j}{{\hat l}_{j + 1}}} } \right)$, where $l_i$ denotes the number of neurons within the actor network's $i$-th layer while ${\hat l}_j$ is the number of neurons within the critic network's $j$-th layer. $I$ and $J$ represent the quantities of fully connected layers for the actor and critic networks, respectively. When PER is introduced, the experience replay complexity increases due to the additional operations required for managing and sampling experiences according to their priorities. Using a SumTree data structure, the time complexity is ${\cal O}\left( {{N_b}\log {N_D}} \right)$. Moreover, the complexity for obtain the optimal beamforming of the BS is $O\left( K \right)$.  Therefore, the time complexity for all $N_e$ episodes can be represented as ${\cal O}\left( {{N_e}{N_s}\left( {\sum\limits_{i = 0}^{I - 1} {{l_i}{l_{i + 1}}}  + \sum\limits_{j = 0}^{J - 1} {{{\hat l}_j}{{\hat l}_{j + 1}}}  + {N_b}\log {N_D} + K} \right)} \right)$.

	\section{Simulation Results}
	\begin{table*}[ht]
		\centering
		\caption{\scriptsize {SYSTEM PARAMETER}}
		\begin{tabular}{|c|c|c|c|c|c|}
			%\hline
			%Parameter & Value & Parameter & Value & Parameter & Value \\
			\hline
			Aircraft mass $m$ & $3$ & Transmission power of BS $P_{\rm BS}^{\rm max}$ & $20$ & Safety variation for roll angle ${\tilde \phi _{\max }}$ & $\pi /12$ \\
			\hline
			Acceleration of gravity $g$ & $9.81$ & Duration of flight $T$ & $30$ & Safety variation for yaw angle ${\tilde \theta _{\max }}$ & $\pi /12$ \\
			\hline
			Number of GUs $K$ & $8$ & Number of time slots $L$ & 60 & Safety variation for pitch angle ${\tilde \psi _{\max }}$ & $\pi /12$ \\
			\hline
			Number of ARIS's elements $N$ & $40$ & Thrust coefficient $C_t$ & $4.848 \times {10^{ - 5}}$ & Nominal no-load motor constant $K_v$ & $380$\\
			\hline
			Number of BS's antennas $M$ & $8$ & Torque coefficient $C_m$ & $8.891 \times {10^{ - 7}}$ & Pass-loss factor $\rho_0$& $10$ \\
			\hline
			Number of sub-surface's elements $\tilde N$ & 10 & Drag coefficient of x-axis $C_{dx}$ & $0.11$ & Maximum speed $v_{\max}$ & 15 \\
			\hline
			No-load current $I_0$ & $0.3$ & Drag coefficient of y-axis $C_{dy}$ & $0.11$ &Maximum acceleration $a_{\max}$ & 5 \\
			\hline
			No-load voltage $U_0$ & $10$ & Drag coefficient of z-axis $C_{dz}$ & $0.2$ & Frame size  & 0.3 \\
			\hline
			Motor resistance $R_0$ & $0.4$ & Safety margin for roll angle $\phi_{\rm max}$ & $\pi/4$ & pass-loss exponents $\alpha_1 / \alpha_2$  & 2 \\
			\hline
			The altitude of ARIS $H$ & 100  & Safety margin for yaw angle $\theta_{\rm max}$ & $\pi/4$ & Rician factors $K_1,K_2$  & 10 \\
			\hline
		\end{tabular}
	\end{table*}
	
	\begin{table}[htbp]
		\renewcommand{\arraystretch}{1.2} % Adjust row spacing
		\centering
		\caption{\scriptsize {HYPERPARAMETERS OF THE ALGORITHM}}
		\begin{tabular}{|>{\centering\arraybackslash}p{0.32\textwidth}|>{\centering\arraybackslash}p{0.12\textwidth}|}
			\hline
			Parameters & Values \\ \hline
			${\text{Episode length}}$ & $1000$ \\ \hline
			{Maximum steps in each episode} & $1000$ \\ \hline
			Replay buffer size & $5\times 10^5$ \\ \hline
			Learning rate for actor network  & $5 \times 10^{-4}$ \\ \hline
			Learning rate for critic network  & $5 \times 10^{-4}$ \\ \hline
			$\text{Discount factor }$ & $0.99$ \\ \hline
			%The temperature learning rate  & $3 \times 10^{-4}$ \\ \hline
			Batch size & $256$ \\ \hline
			%$\text{Optimizer }$ & Adam \\ \hline
		\end{tabular}
		\label{tab:hyperparameters}
	\end{table}
	
	This section provides a comprehensive evaluation of our proposed algorithm for ARIS-assisted communications in terms of the sum-rate. For comparison, the following benchmark schemes are used:
	\begin{itemize}
		\item \textbf{SAC scheme:} We utilize this algorithm to solve the formulated sum-rate maximization problem, which serves as a benchmark to show the superior training efficiency of the PER.
		\item \textbf{PPO scheme:} This method is a popular and reliable DRL algorithm that uses a stochastic policy, which defines a distribution over actions instead of providing a deterministic policy. PPO utilizes a clipped objective function to ensure stable updates, effectively mitigating abrupt policy changes and enhancing training robustness \cite{10049119}.
		\item \textbf{DDPG scheme:} This algorithm integrates deep learning with deterministic policy approaches, designed to handle scenarios characterized by high-dimensional state and continuous action spaces \cite{10704600}. 
		\item \textbf{Fixed RIS scheme:} In this scheme, the ARIS is fixed at $(60,60,H) \text{ m}$, where is the center of the GUs.  Algorithm 1 and Algorithm 2 are performed for the joint optimization of ARIS phase shifts and beamforming at the BS, aiming to show the advantage of flexible deployment of the ARIS.
		%本方案中RIS位置不发生移动，旨在展现ARIS灵活部署的优点
		\item \textbf{Random phase shift scheme:} In this scheme, Algorithm 1 and Algorithm 2 are used to jointly optimize ARIS's altitude, trajectory, and BS beamforming, while the phase shifts of each ARIS sub-surfaces are randomly generated.
		%在该对比方案中，UAV同样采用所提基于欧拉角的控制方法实现轨迹优化，但是所搭载的RIS姿态不发生变化，一直保持水平。
		\item \textbf{ARIS without tilting scheme:} In this comparative baseline scheme, the UAV employs the proposed Euler-angle-based control method for trajectory optimization, while the onboard RIS maintains a fixed horizontal orientation without angular variation \cite{10287195}.
		\item \textbf{Ignoring tilt scheme:} In this scheme, the impact of altitude variations is ignored, but the altitude of ARIS still varies during flight.
	\end{itemize}
	\subsection{Simulation Setup}
	In the simulation, the ARIS is initially positioned at (20, 20, 100) m, while the BS is located at (100, 100, 10) m. The ARIS flies within a $150 \text{ m}\times 150 \text{ m}$ horizontal area bounded by the lower-left corner ${\bf q}_l=(0,0,100)$ m and upper-right corner ${\bf q}_l=(150,150,100)$ m, with its altitude maintained at $100$ m. GUs are randomly distributed across this area. Table II documents the system configurations \cite{9967954,6220873}, while Table III lists the proposed SAC-PER hyperparameter settings, both serving as baseline configurations unless specified otherwise.
	%Unless otherwise specified, the experimental parameters in our simulation and the hyperparameters for the SAC-PER algorithm are depicted in Table II and Table III.
	
	\subsection{Performance Evaluation}
	\begin{figure}[t]
		\centering
		\captionsetup{justification=raggedright, singlelinecheck=false}
		\begin{subfigure}[t]{0.45\textwidth}
			\centerline{\includegraphics[width=\textwidth]{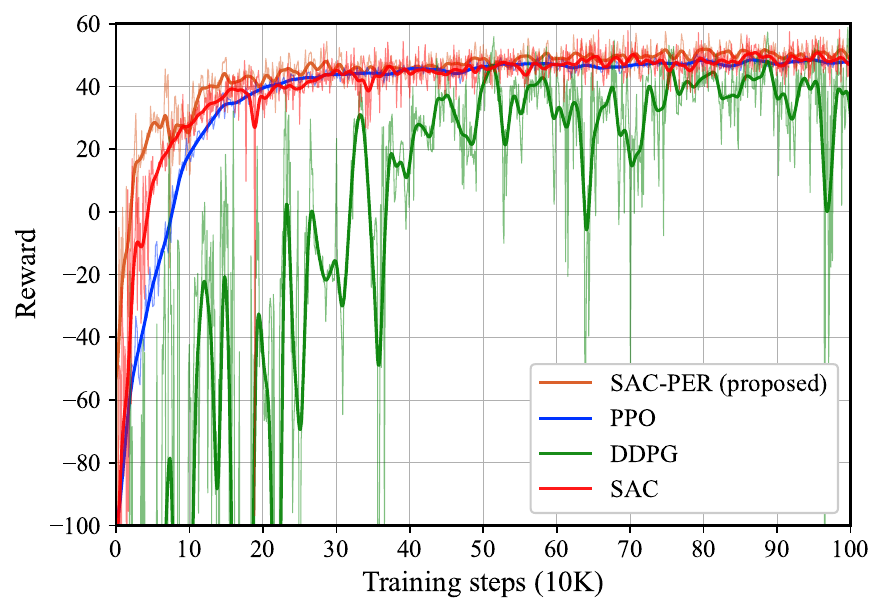}}
			\caption{The convergence performance of different algorithms.}
			\label{fig:convergence}
		\end{subfigure}
		\begin{subfigure}[t]{0.45\textwidth}
			\centerline{\includegraphics[width=\textwidth]{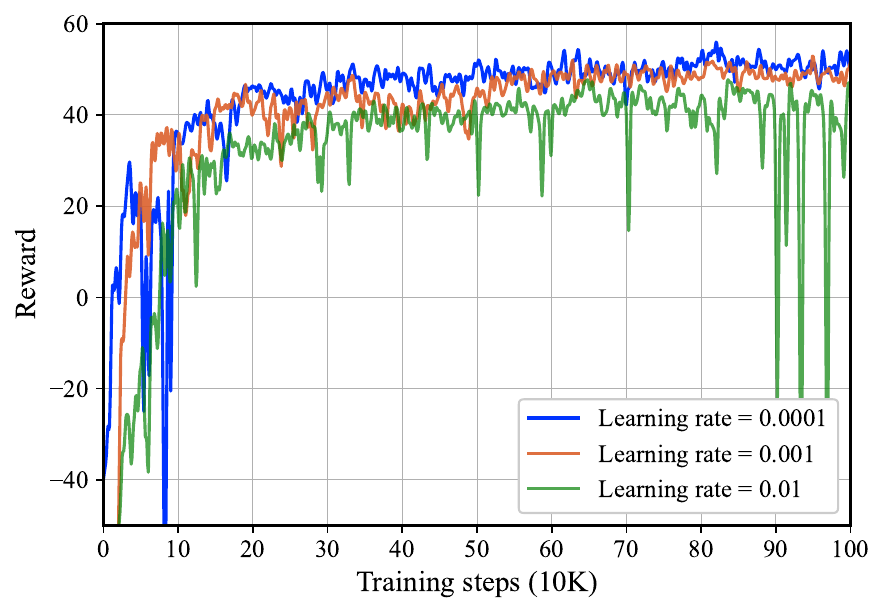}}
			\caption{The performance of the proposed SAC-PER algorithm under different learning rates.}
			\label{fig:convergence_lr}
		\end{subfigure}
		\begin{subfigure}[t]{0.45\textwidth}
			\centerline{\includegraphics[width=\textwidth]{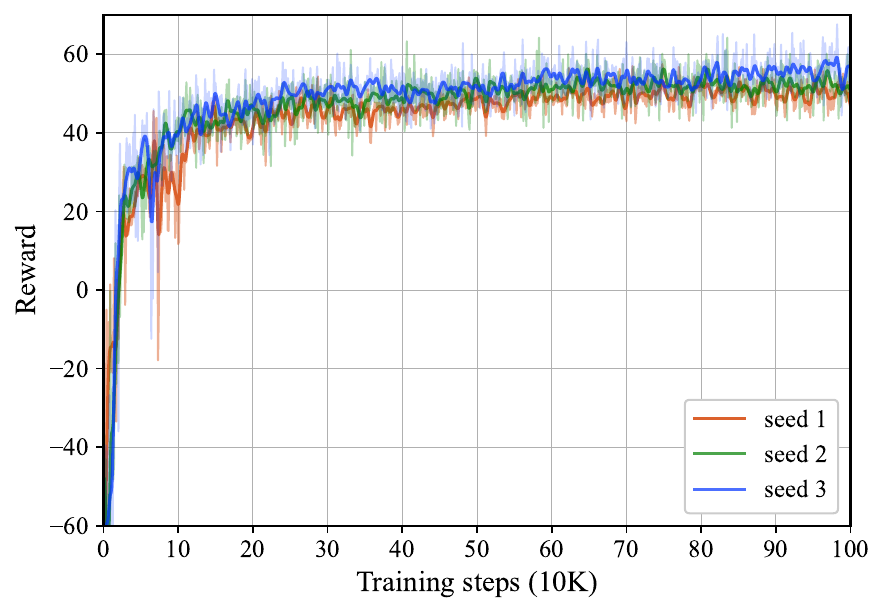}}
			\caption{The performance of the proposed SAC-PER algorithm under different seeds.}
			\label{fig:ES_convergence_seed}
		\end{subfigure}
		\caption{The performance of proposed algorithm.}
	\end{figure}

	\begin{figure}[htbp]
		\centering
		\begin{subfigure}[htbp]{0.24\textwidth}
			\centerline{\includegraphics[width=\textwidth]{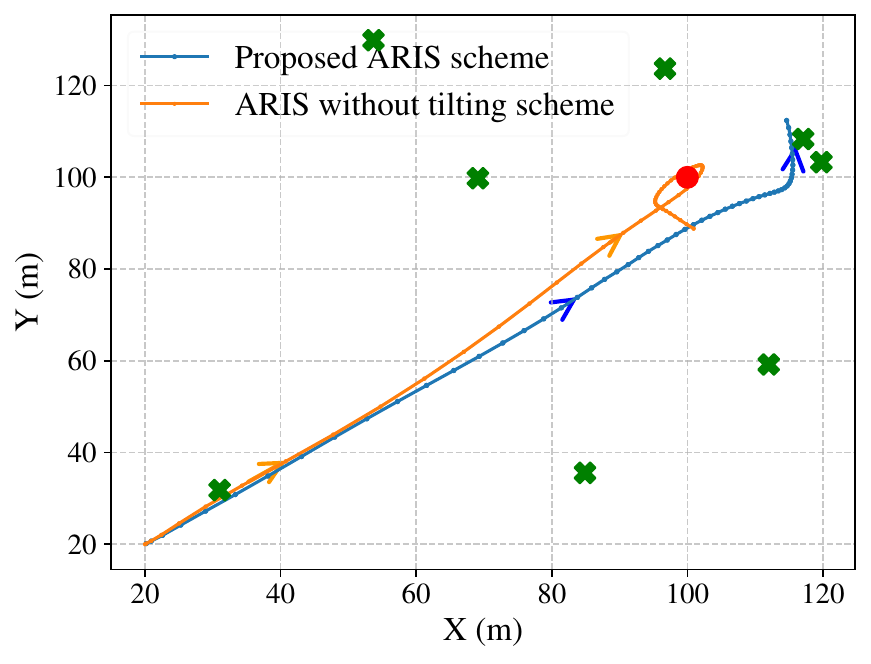}}
			\caption{$H=100$ m, seed=2019.}
			\label{fig:tra1}
		\end{subfigure}
		\begin{subfigure}[htbp]{0.24\textwidth}
			\centerline{\includegraphics[width=\textwidth]{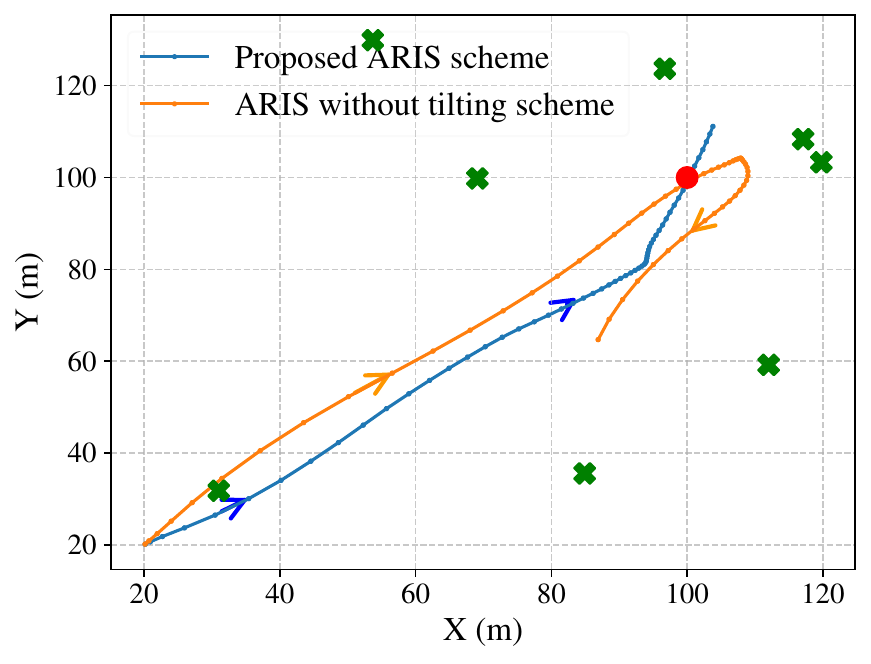}}
			\caption{$H=150$ m, seed=2019.}
			\label{fig:tra2}
		\end{subfigure}
		\begin{subfigure}[htbp]{0.24\textwidth}
			\centerline{\includegraphics[width=\textwidth]{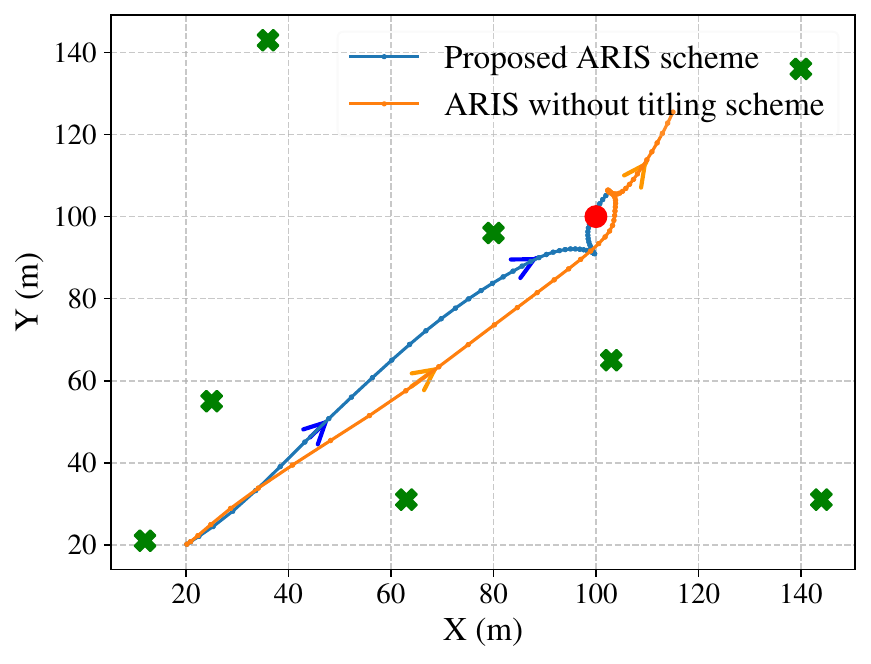}}
			\caption{$H=100$ m, seed=2020.}
			\label{fig:tra3}
		\end{subfigure}
		\begin{subfigure}[htbp]{0.24\textwidth}
			\centerline{\includegraphics[width=\textwidth]{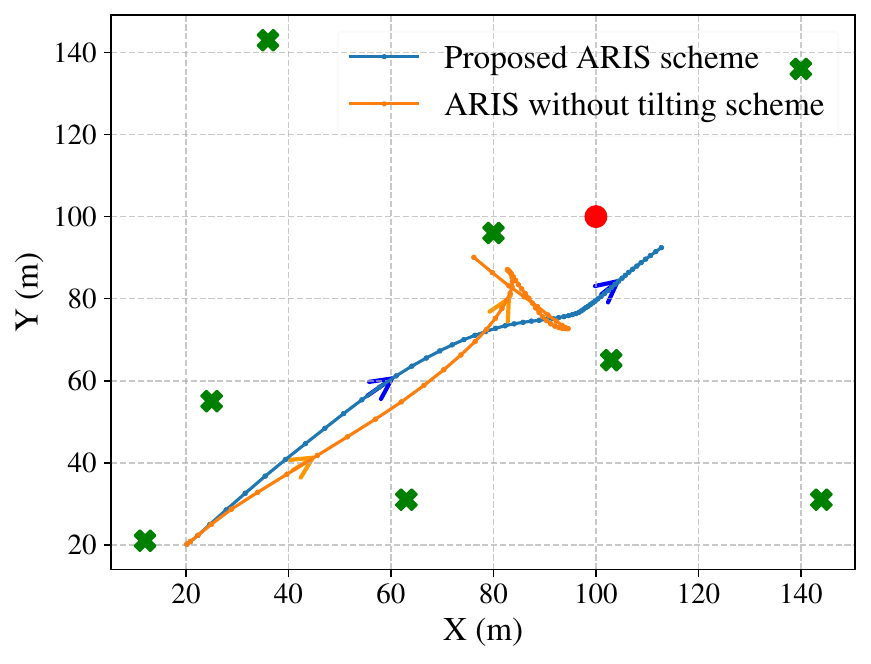}}
			\caption{$H=150$ m, seed=2020.}
			\label{fig:tra4}
		\end{subfigure}
		\caption{The trajectories of ARIS for different random seed and height, where $K=8$.}
	\end{figure}
	\subsubsection{Convergence}
	To verify the effectiveness of the proposed SAC-PER algorithm, we compare it against the SAC, PPO, and DDPG algorithms in Fig. 4(a). As observed, the proposed algorithm achieves faster convergence and superior overall performance compared to the benchmark algorithms. Specifically, SAC-PER converges at around 150K steps, whereas PPO and SAC require approximately 200K steps, and DDPG fails to achieve satisfactory convergence during the entire training process. Furthermore, upon convergence, SAC-PER achieves a significantly higher reward than that of the PPO, highlighting its superior learning efficiency.

	The selection of network parameters plays an important role in DRL. For example, the learning rate significantly affects convergence and network stability. 
	By choosing the appropriate learning rate, the DRL can quickly achieve the desired results. We analyze the impact of learning rate on the SAC-PER algorithm as shown in Fig. 4(b), where the learning rates are set to 0.0001, 0.001, and 0.01, respectively. It can be observed that the best performance is achieved when the learning rate is set to 0.0001, compared to other values. When it is equal to 0.01, the convergence is slow, and it is difficult to converge to a satisfying value, as a large learning rate may cause the step size of each parameter update to be excessively large, resulting in oscillations and instability during the training.
	
	Furthermore, Fig. 4(c) portrays the performance of our proposed SAC-PER algorithm under various random seeds. As observed, the proposed algorithm consistently achieves favorable outcomes across different seeds, which further confirms the applicability of the proposed algorithm for various scenarios.

	\begin{figure}[t]
		\centering
		\includegraphics[width=0.9\columnwidth]{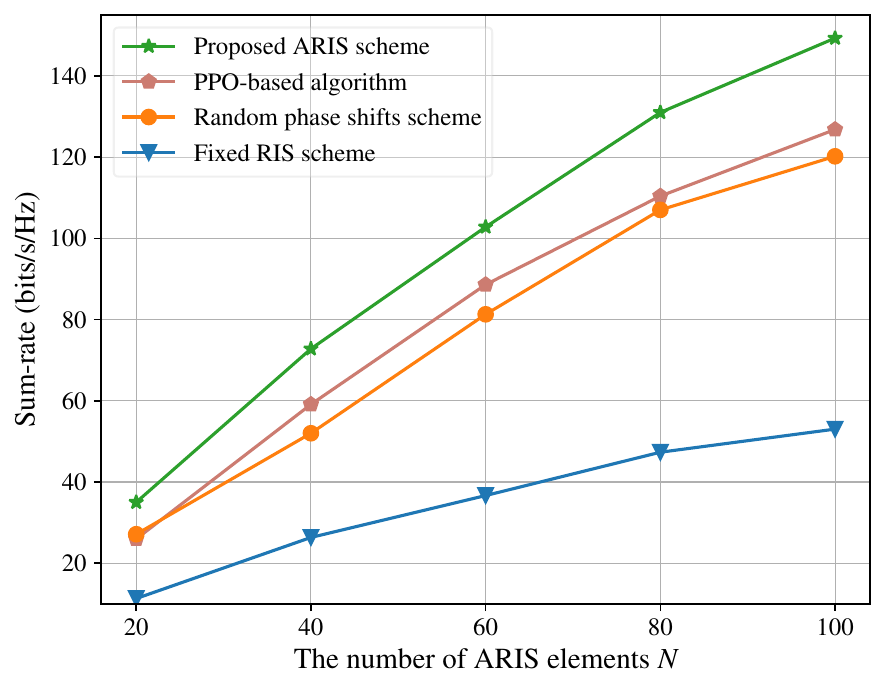}
		\caption{The sum-rate versus the number of ARIS elements $N$, where $M=8$, $K=8$.}
		\label{fig:element}
	\end{figure}
	
	\begin{figure}[t]
		\centering
		\captionsetup{justification=raggedright, singlelinecheck=false}
		\begin{subfigure}[t]{0.45\textwidth}
			\centerline{\includegraphics[width=\textwidth]{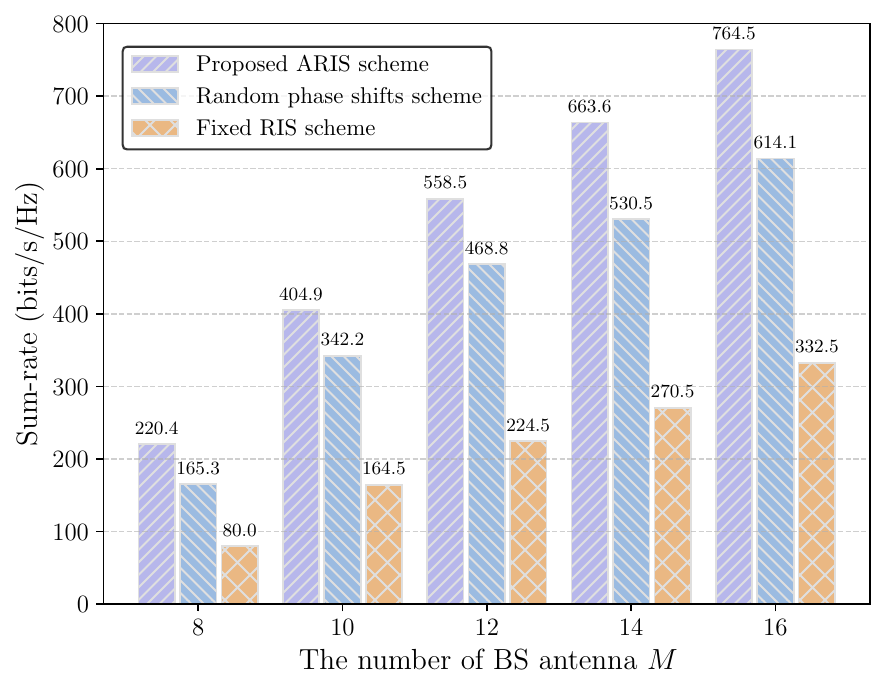}}
			\caption{The sum-rate versus the number of antennas $M$, where $N=200$.}
			\label{fig:antenna}
		\end{subfigure}
		\begin{subfigure}[t]{0.45\textwidth}
			\centerline{\includegraphics[width=\textwidth]{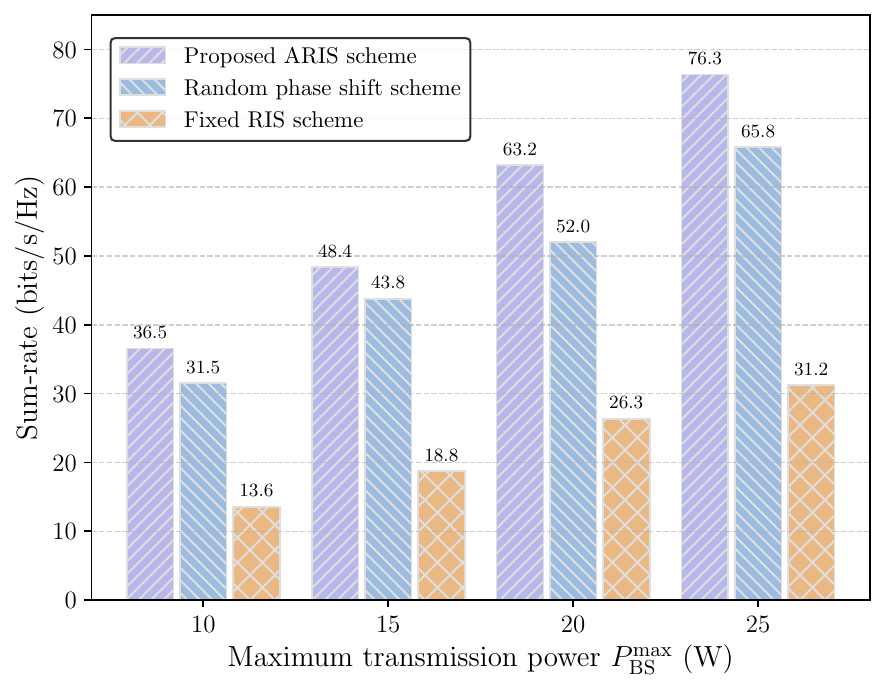}}
			\caption{The sum-rate versus the maximum power of BS $P_{\rm BS}^{\rm max}$, where $M=8$, $N=40$.}
			\label{fig:BS_power}
		\end{subfigure}
		\caption{The sum-rate versus BS antennas and transmission power, where $K=8$.}
	\end{figure}

	\subsubsection{Trajectory}
	Fig. 5 compares the trajectories of the proposed ARIS scheme with the benchmark where the RIS maintains a fixed horizontal orientation. It can be observed that the proposed scheme's trajectories deviate more flexibly to maintain favorable alignment with both the BS and GUs under different seed and height. By contrast, the baseline scheme follows a comparatively rigid path, as it does not adapt its orientation to compensate for changing ARIS's altitude. Specifically, in Figs. 5(a) and 5(c), when $H$ = 100 m, the proposed scheme yields a slightly modified yet more targeted flight trajectory that maintains strong communication links with intermediate GUs. Meanwhile, the baseline scheme, due to the horizontal orientation, occasionally takes a less efficient trajectory in terms of balancing the distances to multiple GUs. Similar trends appear in Figs. 5(b) and 5(d), where the proposed scheme more effectively maneuvers toward areas of higher GUs density and better overall channel quality. Hence, allowing the ARIS to adjust its altitude can lead to improved spatial coverage and greater flexibility compared with the baseline schemes.

	\subsubsection{Sum-rate and RIS elements}
	As shown in Fig. 6, the sum-rate steadily increases with the number of ARIS elements $N$. This trend is intuitive, as a large $N$ provides greater beamforming flexibility, enabling stronger desired signals and more effective suppression of multi-user interference. Moreover, the proposed ARIS scheme outperforms both random phase shifts and fixed RIS schemes in terms of sum-rate, indicating its effectiveness for ARIS's altitude, trajectory, and phase shifts optimization. Moreover, compared to the benchmark PPO scheme, our proposed SAC-PER algorithm could attain greater sum-rate, up to 14.4\%, further demonstrating the advantage of the proposed SAC-PER algorithm in exploration.
	
	\begin{figure}[t]
		\centering
		\includegraphics[width=0.9\columnwidth]{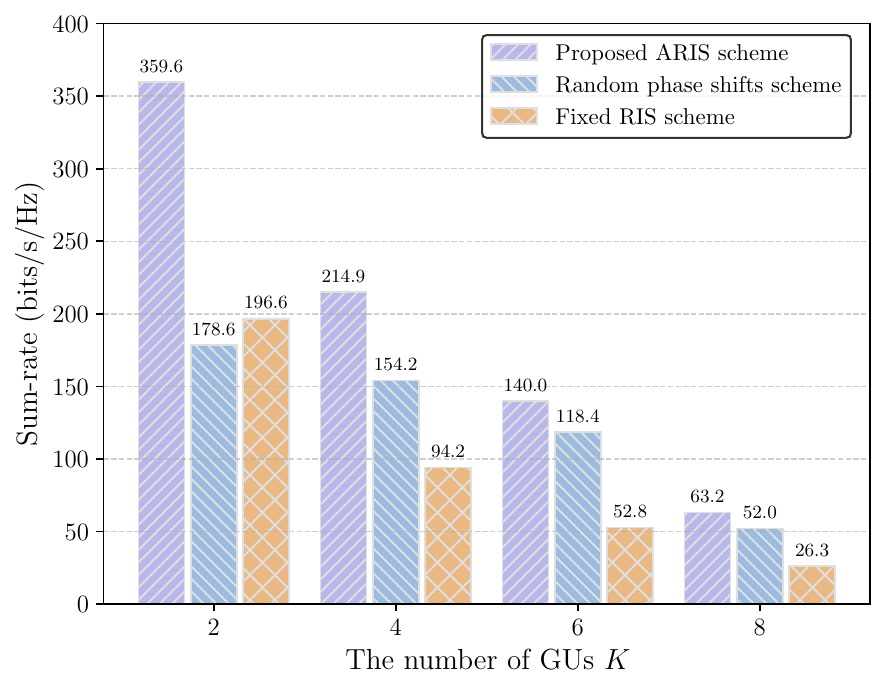}
		\caption{The sum-rate versus the number of GUs $K$, where $N=40$, $M=8$, seed=2019.}
		\label{fig:GUs_number}
	\end{figure}
	
	\begin{figure}[t]
		\centering
		\includegraphics[width=0.45\textwidth]{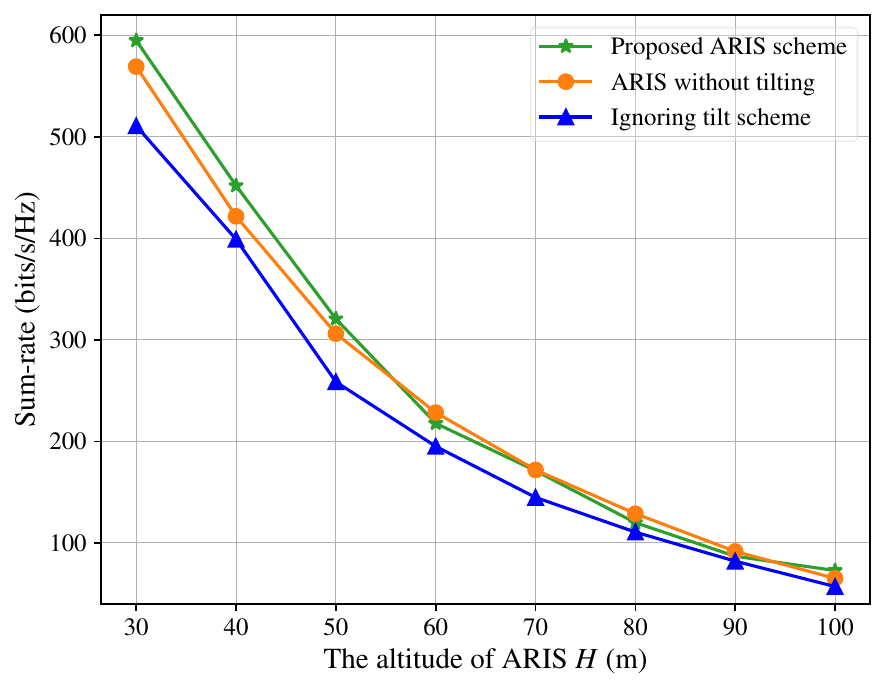}
		\caption{The sum-rate versus the altitude of ARIS, where $N=40$, $M=8$, seed=2018.}
		\label{fig:height}
	\end{figure}
	\begin{figure}[t]
		\centering
		
		\begin{subfigure}[t]{0.24\textwidth}
			\centerline{\includegraphics[width=\textwidth]{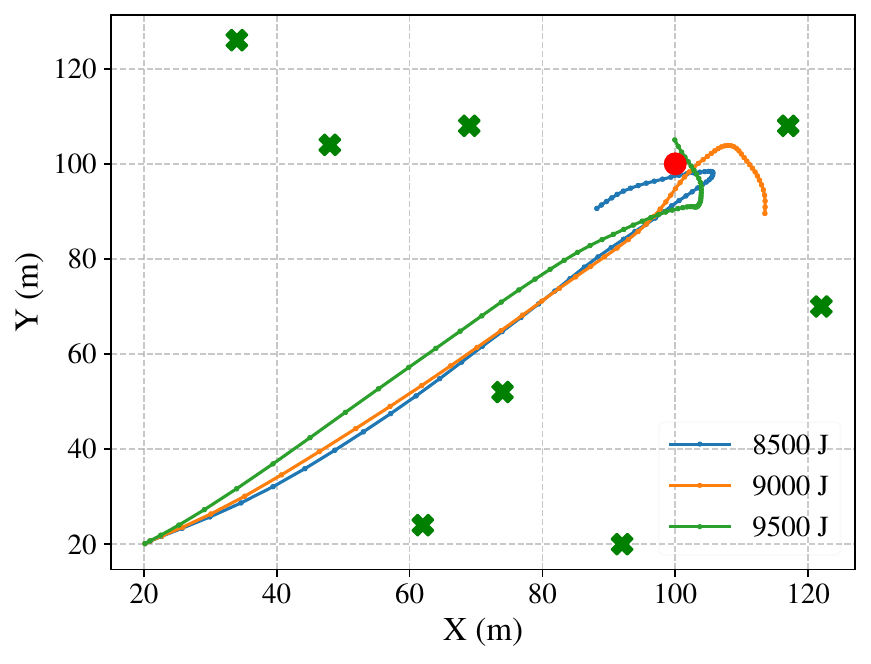}}
			\caption{Trajectory versus flight energy.}
			\label{fig:trace_diff_energy}
		\end{subfigure}
		\begin{subfigure}[t]{0.24\textwidth}
			\centerline{\includegraphics[width=\textwidth]{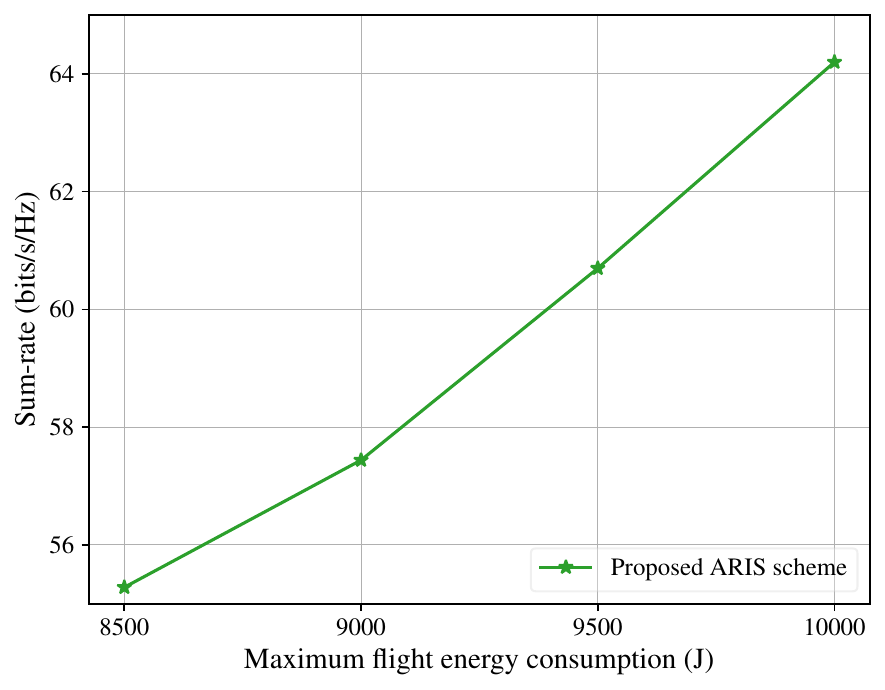}}
			\caption{Sum-rate versus flight energy.}
			\label{fig:fig_diff_energy}
		\end{subfigure}
		\caption{The performance of the proposed scheme under different flight energies.}
	\end{figure}

	\subsubsection{Sum-rate and antennas}
	In Fig. 7, we evaluate the performance gain of the proposed ARIS scheme under different numbers of BS antennas. Specifically, we set the number of GUs $K=8$. As illustrated in Fig. 7(a), where the number of ARIS element $N=200$, the sum-rate grows as the BS antenna count increases. This is because, under ZF precoding, more antennas provide greater spatial degrees of freedom and stronger interference-cancellation capability, thereby improving the overall channel gain. Furthermore, in multi-user systems, a larger number of antennas can better allocate beams to each GU, reducing interference and ultimately enhancing system capacity. Furthermore, we analyze the sum‐rate of different schemes under identical number of BS's antennas but varying maximum transmission power in Fig. 7(b). It is evident that, as the transmission power rises, the performance gain of our proposed scheme becomes increasingly prominent relative to the benchmark schemes. This finding highlights the effectiveness of jointly optimizing the ARIS trajectory, altitude, and phase shifts in further enhancing the overall spectral efficiency, particularly when sufficient transmission power is available.

	\begin{figure}[t]
		\centering
		\includegraphics[width=0.8\columnwidth]{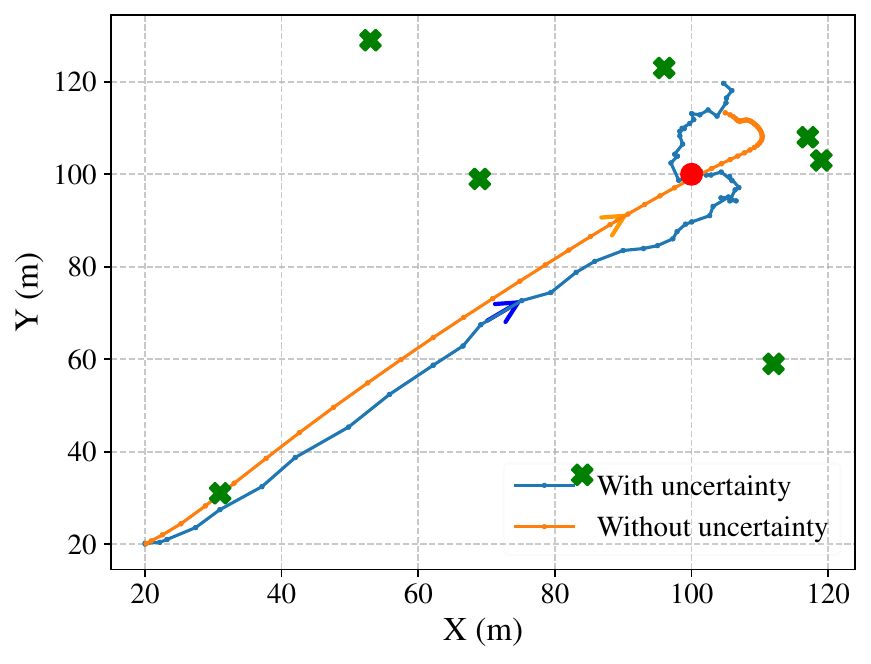}
		\caption{The trajectory for the proposed ARIS scheme with uncertainty.}
		\label{fig:trace_robust}
	\end{figure}
	\begin{figure}[t]
		\centering
		
		\begin{subfigure}[htbp]{0.24\textwidth}
			\centerline{\includegraphics[width=\textwidth]{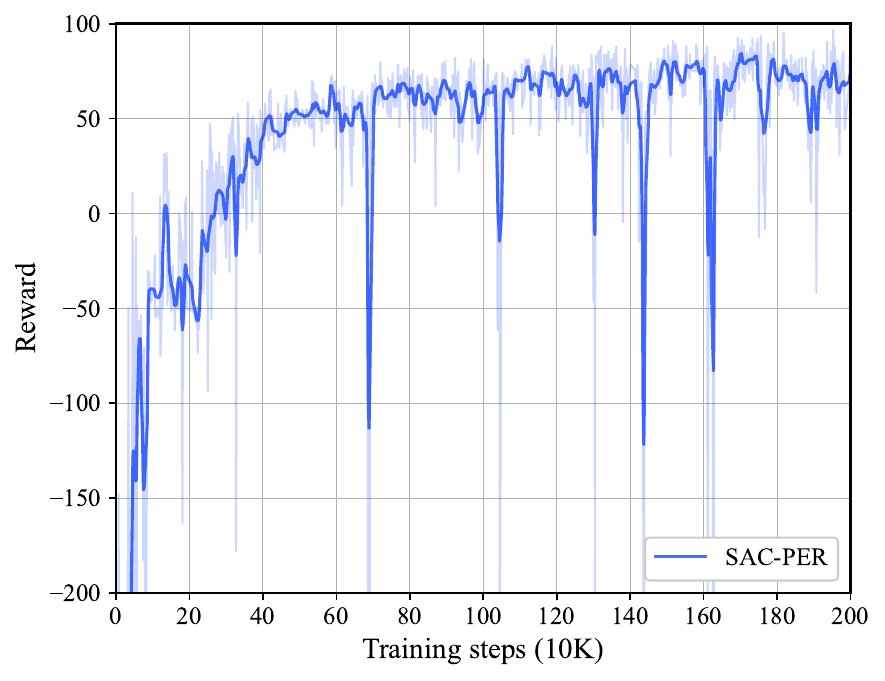}}
			\caption{The convergence performance.}
			\label{fig:convergence_multi_UAV}
		\end{subfigure}
		\begin{subfigure}[htbp]{0.24\textwidth}
			\centerline{\includegraphics[width=\textwidth]{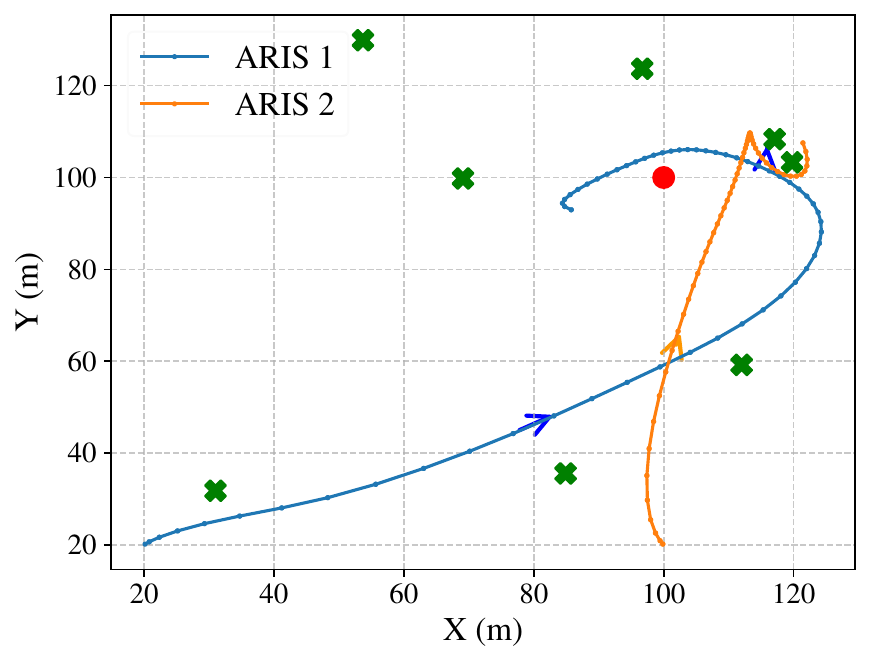}}
			\caption{The trajectories of ARISs.}
			\label{fig:trace_multi_UAV}
		\end{subfigure}
		\caption{The performance of the proposed scheme for multi-ARIS, where $I=2$.}
	\end{figure}
	\subsubsection{Sum-rate and GUs}
	To demonstrate the extensibility of the proposed scheme, we compare the performance of different approaches under different numbers of GUs $K$. Given the random nature of GUs positions, we fix the random seed for all schemes to 2019, ensuring consistency in GUs distribution at each GUs count. As shown in Fig. 8, when $K$ is relatively small, all three schemes achieve their highest sum‐rate. This is because, in the low‐GUs regime, the ZF can fully exploit the available spatial degrees of freedom, effectively mitigating multi-user interference to a negligible level. Moreover, in all tested scenarios, the proposed scheme consistently outperforms the baseline methods, further highlighting its superiority.
	
	\subsubsection{Sum-rate and ARIS altitude}
	%To clearly compare the performance differences between the proposed ARIS scheme and the conventional ARIS without tilting scheme, we further analyze the total rate of both schemes at different flight altitudes. As depicted in Fig. 9, the proposed scheme demonstrates significant advantages when the ARIS flight altitude is below 50 m. This phenomenon can be attributed to the pronounced dynamic impact of ARIS altitude variations on signal incidence and reflection angles in low-altitude scenarios. The baseline schemes, due to its fixed ARIS orientation and oversight of altitude variations, fail to adapt to such dynamic channel variations. In contrast, the proposed ARIS scheme employs an Euler angles-based control framework to dynamically adjust the spatial orientation of reflecting elements. Combined with the SAC-PER algorithm for joint optimization of altitude, trajectory, and phase shifts, it actively compensates for signal path deviations caused by altitude perturbations, thereby enhancing channel gain. This approach effectively balances multi-user coverage and signal enhancement requirements. These results validate the technical superiority of the altitude-integrated ARIS scheme in dynamic environments and provide theoretical support for the engineering deployment of ARIS systems.
	To clearly compare the performance differences among different ARIS control strategies, we further evaluate the sum-rate of three schemes at various flight altitudes, as illustrated in Fig.~9. The proposed ARIS scheme is compared with two benchmark schemes, namely the ARIS without tilting scheme which maintains a fixed ARIS orientation, and the ignoring tilt scheme which allows altitude variations during flight but neglects the effect of ARIS attitude variations. The proposed scheme exhibits significant advantages, especially at low altitudes, where the impact of altitude-induced angular deviations on signal incidence and reflection is more pronounced. In contrast, the baseline schemes, due to their lack of dynamic attitude adjustment or omission of tilt effects, fail to adapt to such variations and suffer from degraded channel alignment. By integrating an Euler angles-based control mechanism with the SAC-PER algorithm, the proposed scheme jointly optimizes the ARIS’s altitude, trajectory, and phase shifts. This allows the ARIS elements to dynamically align with the optimal signal reflection directions, thereby compensating for misalignment caused by flight perturbations and improving overall channel gain. As a result, the proposed method effectively balances multi-user coverage with directional signal enhancement. These results validate the technical superiority of the altitude-integrated ARIS scheme in dynamic environments and provide theoretical support for the engineering deployment of ARIS systems.

	\subsubsection{Sum-rate and flight energy}
	To demonstrate the influence of the flight energy consumption, we compare the ARIS's trajectory and sum-rate under different flight energy budget.
	As illustrated in Fig.~10(a), the ARIS trajectories exhibit significant variations under three different maximum flight energy constraints, namely 8500~J, 9000~J, and 9500~J. Under the 9500~J energy budget, the ARIS demonstrates more aggressive motion behavior in the initial time slots, characterized by higher acceleration and longer displacement per time slot. This phenomenon arises because, according to equations~(7) and~(8), a higher acceleration requires larger roll and pitch angles, which, based on equation~(13), results in greater flight energy consumption. With a larger power budget, the ARIS is capable of executing more rapid maneuvers despite the higher energy cost. Furthermore, as depicted in Fig.~10(b), the sum-rate increases with the available flight energy. This is attributed to the ARIS reaching favorable positions with higher channel gains more quickly, thereby enhancing the overall communication performance.

	\subsubsection{Robustness analysis}
		In practical scenarios, due to inaccurate positioning information, wind gusts, and other factors, the ARIS may deviate from the scheduled trajectory, which may affect the communication performance. Therefore, in order to adapt to actual scenarios, the unpredictable ARIS trajectory caused by uncertainties should be specially addressed to design a robust ARIS-assisted communications. The uncertainty trajectory can be modeled as 
		\begin{equation}
			{\hat{\bf{q}}}[l] = {\bf{q}}[l] + \Delta {\bf{q}}[l],\;\forall l \in {\cal L},
		\end{equation}
		where ${\bf{q}}[l]$ is the scheduled trajectory and $\Delta {\bf{q}}[l]$ is the position error caused by uncertainties. According to \cite{10068215}, the uncertainty can be modeled as a Gaussian random variable, given by
		\begin{equation}
			\Delta {\bf{q}}[l] \sim {\cal N}\left( {0,\varepsilon _0^2{\bf I}} \right),\;\forall l \in {\cal L},
		\end{equation}
		where {\bf I} is a third-order identity matrix corresponding to the three dimensions in space. Note that although we have assumed that the ARIS flight at a fixed height, there are still uncertainties in the vertical dimension. In Fig. 11, we compare the trajectories and it can be seen that the proposed scheme can effectively adapt to the uncertainty caused by factors such as wind gusts.

		\subsubsection{Multi-ARIS scenario}
		Considering that the collaboration between ARISs can further enhance the communication performance and coverage, we further consider the scenario of multi-ARIS-assisted communications. First, we define the set of ARISs as 
		${\cal{I}}=\left\{1,\dots,i,\dots,I \right\}$. The gain from ARIS $i$ to GU $k$ can still be calculated using equation (26), denoted as ${{\boldsymbol{\xi}}_{i,k}}$. Notably, since multi-ARIS is introduced, the concatenated channel ${\bf v}_k$ defined previously would become ${{\bf{v}}_k}[l] = \sum\limits_{i = 1}^I {{\bf{h}}_{i,k}^H[l]{{\boldsymbol{\xi}}_{i,k}}[l]{{\bf{H}}_i}[l]}  + {\bf{h}}_{{\rm{BS}},k}^H$. Furthermore, to ensure safe flight of multi-ARIS, we introduce a minimum distance constraint:
		\begin{equation}
			{\left\| {{{\bf q}_i}[l] - {{\bf q}_j}[l]} \right\|^2} \ge d_{\min }^2,\forall i,j \in {\cal I}, i \ne j, l \in {\cal L}.
		\end{equation}	
		
		We continue to adopt the proposed SAC-PER algorithm to solve this problem. The state space is augmented by incorporating the Euler angles, position, velocity, and remaining flight energy of each ARIS at every time slot. Meanwhile, the action space is extended to include the variations of Euler angles and the phase shifts of each sub-surface. It is worth noting that, due to the introduction of new constraints, the reward function is redesigned to ensure flight safety, given by
		\begin{equation}
			{r_t} = 
			\bar{R}[l] - P_4, \text{if} \; {\left\| {{{\bf{q}}_i}[l] - {{\bf{q}}_j}[l]} \right\|^2} < {d_{\min }^2},\forall i,j \in {\cal I}, i \ne j.
		\end{equation}
		where the penalty $P_4$ is introduced to keep all ARIS at a safe distance.
	
		As illustrated in Fig. 12(a), the proposed SAC-PER algorithm maintains strong performance in the multi-ARIS scenario, achieving convergence within approximately 400K steps. Compared to the single-ARIS-assisted case, it yields improved communication performance. Furthermore, Fig. 12(b) depicts the trajectories of two ARISs, which clearly demonstrate the effectiveness of the proposed algorithm in optimizing the trajectories of multiple ARISs while ensuring flight safety.

	\section{conclusion}
	In this paper, we have investigated an ARIS-assisted wireless communication system, where a quadrotor UAV is equipped with a RIS to enhance signal reflection. Unlike prior works that assume a persistently horizontal RIS, we have incorporated the UAV’s dynamics and developed an Euler-angles-based control framework, enabling simultaneous trajectory and altitude optimization. To maximize the system sum-rate, we have jointly optimized the UAV’s trajectory, RIS phase shifts, and BS beamforming. Given the strong coupling among these variables, the problem was formulated as an MDP, and a deep reinforcement learning algorithm based on SAC-PER was proposed to determine the ARIS’s Euler angles and phase shift. Additionally, the BS beamforming was optimized via a bisection-assisted water-filling algorithm under given actions.
	Simulation results have demonstrated that the proposed algorithm achieves superior communication performance and converges to high-quality solutions. Importantly, the integration of altitude control into trajectory design has provided a more practical and flexible framework for real-world ARIS deployment.
	Beyond performance gains, our findings have highlighted that explicitly considering UAV tilt and altitude variations can fundamentally influence UAV control strategies and RIS configuration. On the control side, adaptive UAV flight strategies must dynamically couple altitude variation and trajectory to maintain beam alignment under realistic disturbances. On the RIS side, the configuration should be co-designed with UAV dynamics to achieve stable performance in fluctuating environments. These were often overlooked in conventional ARIS-assisted models. Future research could extend this framework to more challenging settings, including dynamic user mobility, imperfect CSI, and distributed multi-agent learning frameworks.

	\bibliographystyle{IEEEtran}
	\bibliography{references.bib}  % references.bib 是你的 BibTeX 文件名
\end{document}